\documentclass[sn-mathphys,Numbered]{sn-jnl} 
\usepackage{graphicx}%
\usepackage{multirow}%
\usepackage{amsmath,amssymb,amsfonts}%
\usepackage{amsthm}%
\usepackage{mathrsfs}%
\usepackage[title]{appendix}%
\usepackage{xcolor}%
\usepackage{textcomp}%
\usepackage{manyfoot}%
\usepackage{booktabs}%
\usepackage{algorithm}%
\usepackage{algorithmicx}%
\usepackage{algpseudocode}%
\usepackage{listings}%
\usepackage{verbatim}%
\usepackage{tabularx}

\newcommand{\kms}{km\,s$^{-1}$}

\begin{document}

\title[Diagnostics of the solar coronal plasmas by magnetohydrodynamic waves]{Diagnostics of the solar coronal plasmas by magnetohydrodynamic waves: Magnetohydrodynamic seismology}


\author*[1,2]{\fnm{Valery M.} \sur{Nakariakov}}\email{V.Nakariakov@warwick.ac.uk}
\author[1]{\fnm{Sihui} \sur{Zhong}}
\author[1,2]{\fnm{Dmitrii Y.} \sur{Kolotkov}}
\author[1]{\fnm{Rebecca L.} \sur{Meadowcroft}}
\author[1]{\fnm{Yu} \sur{Zhong}}
\author[3,4]{\fnm{Ding} \sur{Yuan}}


\affil*[1]{\orgdiv{Centre for Fusion, Space \& Astrophysics, Department of Physics}, \orgname{University of Warwick}, \orgaddress{
\city{Coventry} \postcode{CV4 7AL}, \country{United Kingdom}}}
\affil[2] {\orgdiv{Engineering Research Institute \lq\lq Ventspils International Radio Astronomy Centre (VIRAC)\rq\rq}, \orgname{Ventspils University of Applied Sciences}, \orgaddress{\city{Ventspils}, \postcode{LV-3601}, \country{Latvia}}}
\affil[3]{\orgdiv{Shenzhen Key Laboratory of Numerical Prediction for Space Storm}, \orgname{Harbin Institute of Technology, Shenzhen}, \orgaddress{\postcode{518055}, \state{Guangdong}, \country{China}}}
\affil[4]{\orgdiv{Key Laboratory of Solar Activity and Space Weather}, \orgname{National Space Science Center, Chinese Academy of Sciences}, \orgaddress{\city{Beijing}, \postcode{100190}, \country{China}}}


\abstract{Macroscopic wave and oscillatory phenomena ubiquitously detected in the plasma of the corona of the Sun are interpreted in terms of magnetohydrodynamic theory. Fast and slow magnetoacoustic waves are clearly distinguished in observations.
Properties of coronal magnetohydrodynamic waves are determined by local parameters of the plasma, including the field-aligned filamentation typical for the corona. It makes coronal magnetohydrodynamic waves reliable probes of the coronal plasma structures by the method of magnetohydrodynamic seismology. 
For example, propagating slow waves indicate the local direction of the guiding magnetic field. Standing, sloshing and propagating slow waves can be used for probing the coronal heating function and the polytropic index. 
Kink oscillations of coronal plasma loops provide us with the estimations of the absolute value of the magnetic field in oscillating plasma loops. This tutorial introduces several techniques of magnetohydrodynamic seismology of solar coronal plasmas. It includes the description of practical steps in the data acquisition, pre-processing, and processing using the open-access data of the Atmospheric Imaging Assembly on the Solar Dynamics Observatory spacecraft, and elaborated data analysis techniques of motion magnification and Bayesian statistics.  
}



\maketitle

\section{Introduction}\label{sec1}

The corona of the Sun is the outermost, almost fully ionised part of the solar atmosphere (e.g., \citep{2019ARA&A..57..157C}). 
The corona is penetrated by the magnetic field generated by dynamo processes in the solar interior, and emerging through the solar surface in a number of concentrated magnetic elements, pores and sunspots. In the corona, the magnetic field forms so-called closed magnetic structures when the field lines are bent down to the surface of the Sun, i.e., remain within the visible part of the corona. Once reaching the corona, the magnetic field spreads out, filling up all the volume of this part of the solar atmosphere. Some parts of the closed corona, usually above major sources of the surface magnetic field, such as pores and sunspots, appear to be highly dynamic and host various eruptions, mass ejections and flares. These regions are known as active regions. The closed corona outside active regions is \lq\lq diffuse\rq\rq. Regions in which the magnetic flux appears to go out of the Sun towards the heliosphere and beyond constitute the \lq\lq open\rq\rq\ corona. Parts of the open corona which appear to have the plasma with the reduced density are called coronal holes. 

Absolute values of the coronal magnetic field reach several hundred Gauss above pores and sunspots, and are typically several tens of Gauss in active regions, decreasing to a few Gauss or lower in the diffuse corona and coronal holes. The field decreases with height. The topology of the coronal magnetic field is highly complicated, and is very far from being dipolar, with a huge number of elements of the opposite polarity, resolved at the photosphere. The measurement {by the Zeeman effect} of the magnetic field in the corona is usually impossible, and its estimation is often based on the extrapolation of photospheric magnetic sources. In some cases, the field can be estimated by radio observations of the  {free--free}, gyroresonant and gyrosynchrotron emissions,  {see, e.g., \cite{2021FrASS...7...77A}, and references therein. } 

The coronal plasma is mainly hydrogen, with some fraction of alpha particles and various ions of heavier elements, so that the effective mean particle mass in the corona is 0.6--0.7 of the proton mass.
Typical temperatures of the coronal plasma range from several hundred thousand K to a few million K during the quiet periods of the activity. During solar flares, the temperature of the plasma structures involved in flares, for example, in flaring active regions, can reach several tens of million K. 
Typical electron concentrations in the lower part of the corona range from about $10^{8}$~cm$^{-3}$ in coronal holes to about $10^{10}$~cm$^{-3}$ in active regions, reaching $10^{12}$~cm$^{-3}$ in flaring regions. The coronal plasma concentration is subject to gravitational stratification. The stratification scale height is linearly proportional to the temperature, and is about $5 \times 10^9$~cm for the temperature of one million K.
The coronal plasma parameter $\beta$, defined as the ratio of the gas to magnetic pressures, is typically much lower than unity.

The solar corona attracts growing attention for several reasons. From the point of view of practical applications, the corona is the birthplace of drivers of extreme events of space weather, such as flares and coronal mass ejections. These impulsive releases of the coronal magnetic energy can disrupt or damage various technological systems in space, atmosphere and on the ground, and their forecasting is of growing interest. Furthermore, being a natural plasma environment, the corona offers the plasma physics research community a great opportunity to study various physical processes of fundamental importance, such as magnetic reconnection, acceleration of charged particles to relativistic energies, plasma turbulence and micro-turbulence, transport coefficients, magnetohydrodynamic (MHD) waves and jets, and many others. In other words, the corona is a natural plasma physics laboratory. Coronae are known to exist in other cool stars, including sun-like stars which may host habitable exoplanets. Violent dynamic processes in the corona are potentially {dangerous for life on the planets on one hand, and, on the other hand, according to some studies could presumably provide necessary conditions} for the appearance of extraterrestrial life, e.g., \cite{2018SciA....4.3302R}. The solar corona is the only stellar corona open to {high spatial and temporal resolution} observational study. 

\begin{figure}
\centering
\includegraphics[width=0.9\linewidth]{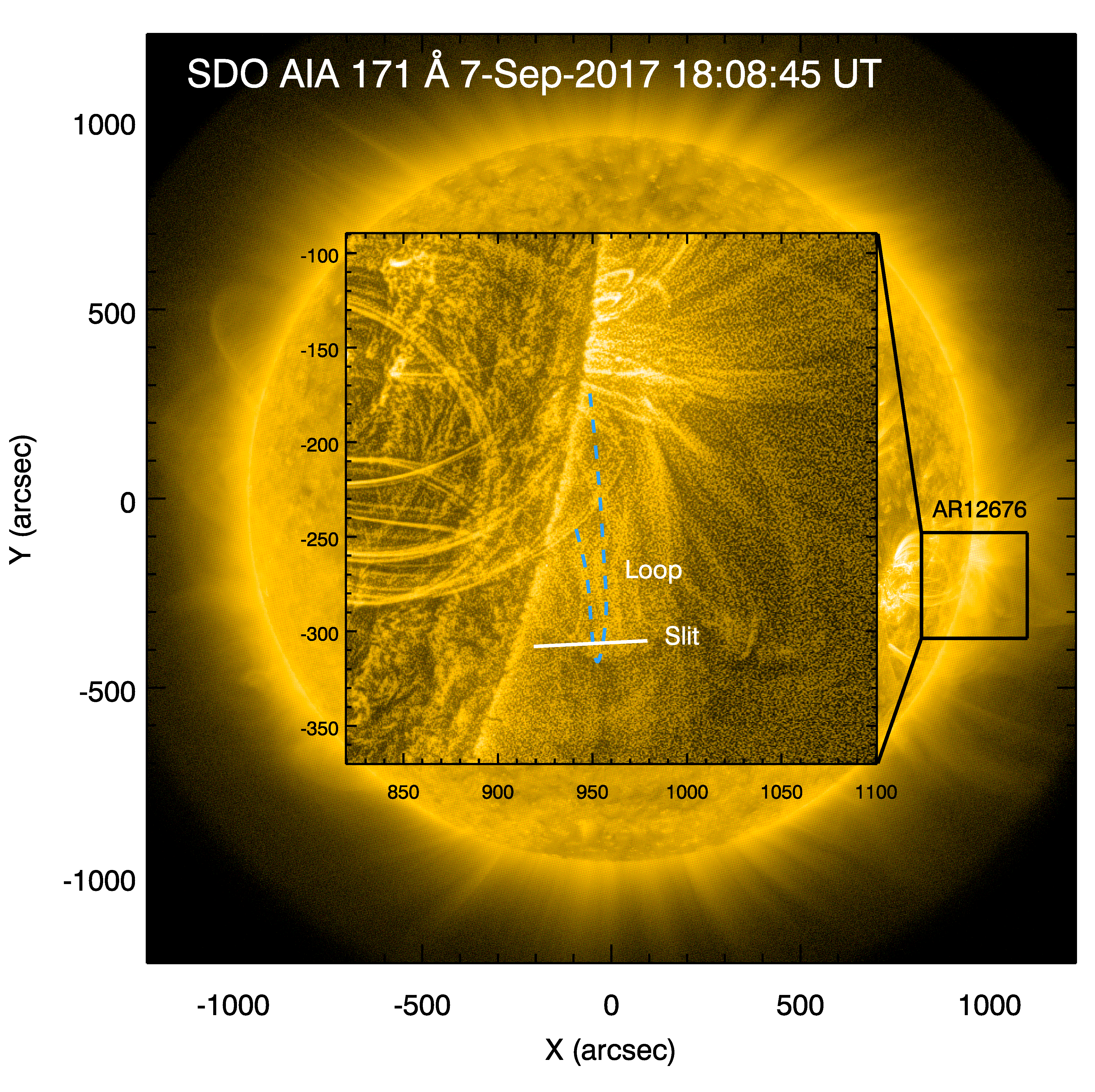} 
\caption{An image of the solar corona taken in the EUV band with the SDO/AIA instrument. In the central box, a zoomed region of interest is shown, indicated by the black lines in the full solar disk image in the periphery.
The brightness of a pixel is proportional to the number of photons with wavelengths in the vicinity of 171\,\AA\ coming to the pixel. This emission is produced by the coronal plasma with a temperature of about 10$^6$~K. {The zoomed box is enhanced using the multiscale Gaussian normalization method to highlight coronal loops \cite{2014SoPh..289.2945M}. The blue curve inside presents a typical plasma loop.}  
\label{fig:cor}
}
\end{figure}

Despite an intensive multi-wavelength study of the corona of the Sun from space and ground, a number of crucial physical parameters of the corona cannot be reliably estimated. In particular, it is commonly accepted that the magnetic field determines the structure, morphology and energetics of the corona. However, standard astronomical techniques for the diagnostics of the magnetic field by, for example, the Zeeman effect, are not applicable to the one-million K plasma of the corona. {Emission lines experience strong thermal broadening at such high temperatures, making any Zeeman splitting challenging to measure and analyse accurately.} Another key parameter of the corona, which makes it very different from a similar and also intensively studied plasma environment, the Earth's magnetosphere, is its fine field-aligned structuring (e.g., \cite{2016SSRv..200...75N}). Figure~\ref{fig:cor} demonstrates a typical image of the solar corona taken in the extreme ultraviolet (EUV) band. The fine filamentation appears as bright, highly elongated structures. The bent bright threads are called coronal plasma loops. 
Typically, the loop's major radii range from a few tens to several hundred thousand km, while the minor radii are a few thousand km. Planes of the loops are usually tilted from the vertical direction. It is believed that a loop represents a magnetic flux tube, i.e., the magnetic field is tangential to its boundary. However, minor radii of loops do not show any significant increase with height (e.g., \citep{2020ApJ...900..167K}), which is a puzzle as one would expect the magnetic flux tube to have a dipolar shape. 
It is also not clear whether characteristic scales of the coronal plasma filamentation, in particular, the minor radii of the loops are fully resolved with available observational instruments. In polar regions where the main part of the magnetic flux is open into the solar wind, there are also bright structures stretched radially outward from the Sun, called polar plumes. 

In almost all observational bands the corona is an optically thin, i.e., transparent medium. The observed intensity of the radiation is the result of the integration of some function of the plasma parameters along the line-of-sight (LoS). For example, for the thermal emission such as in EUV, soft X-ray and thermal microwave bands, it is the integral of the product of the density squared and a function of the plasma temperature, specific for the specific wavelength and the observational instrument.

An effective plasma diagnostics technique is the use of MHD waves which carry information about the medium which supports them. In the corona, this technique is called MHD seismology (e.g., \cite{2014SoPh..289.3233L}). In laboratory plasmas, this approach is known as MHD spectroscopy (e.g., \cite{2001PhLA..289..127S, 2002PPCF...44B.159F}). A similar technique used in magnetospheric research is so-called magnetoseismology, see, e.g., \cite{2005GeoRL..3218108C, 2014JGRA..119.8338T, 2023BAAS...55c.063C} for the Earth' magnetosphere, and \cite{2019JGRA..124..211J} for the magnetosphere of Mercury. MHD seismology of the corona should not be confused with helioseismology (e.g., \cite{2016LRSP...13....2B}) and astroseismology  (e.g., \cite{2013ARA&A..51..353C, 2019LRSP...16....4G}) which diagnose interiors of the Sun and stars by oscillations of the Sun's or star's surface, respectively, {and also seismology of the Earth's interior}. A spin-off branch of coronal seismology is prominence seismology \cite{2009SSRv..149..175O, 2009SSRv..149..283T,2020ApJ...899...99O}, which specifically addresses quiescent and erupting partly-ionised relatively cold prominences and filaments, and shares common philosophy and methodology with the seismology of the warm and hot coronal plasma structures.  

The availability of high-resolution observations simultaneously in the time and spatial domains allows for the detection of coronal MHD waves as propagating or standing displacements of various plasma structures or variations of the EUV, X-ray or microwave emission, or oscillatory and/or propagating Doppler shifts of coronal emission lines.   
In contrast with other plasma environments, such as the solar wind and Earth's and planetary magnetospheres, where \textit{in situ} measurements of magnetic field perturbations are possible, the incompressive nature of Alfv\'en waves prevents their unequivocal identification in imaging and spectral observational data.
However, observational detections of various magnetoacoustic waves in the corona are abundant (see, e.g., \cite{2020ARA&A..58..441N} for a recent comprehensive review). Low values of the plasma parameter $\beta$ in the corona make the Alfv\'en $C_\mathrm{A}$ and sound $C_\mathrm{s}$ speeds to be significantly different from each other as $C_\mathrm{s}/C_\mathrm{A} \approx \beta^{1/2}$. For example, in plasma loops of coronal active regions, typically $C_\mathrm{A}\approx 1,000$~km\,s$^{-1}$ and $C_\mathrm{s}\approx 150$~km\,s$^{-1}$, which allows for the confident discrimination between fast and slow magnetoacoustic waves. 
The key feature of the solar coronal plasma which determines properties of magnetoacoustic waves is its field-aligned filamentation. 

{In this tutorial paper we briefly introduce the theory of MHD waves in a simple, 1D perpendicular plasma non-uniformity (Section~\ref{sec:prop}), summarise main MHD wave phenomena observationally detected in the corona (Section~\ref{sec:obs}), describe practical steps in the acquisition, processing and seismological analysis of observational data, illustrating it on the EUV imager SDO/AIA data which is frequently used in coronal seismology (Section~\ref{sec:data}), give several example recipes of seismological inversions (Section~\ref{sec:example}), and share our thoughts about the future perspectives in coronal MHD seismology in Section~\ref{sec:con}.}

\section{Properties of MHD waves in solar coronal plasma structures}\label{sec:prop}

Typical oscillation periods of magnetoacoustic waves detected in the corona are from a few seconds to several tens of minutes. These values are several orders of magnitude greater than plasma oscillation periods and gyroperiods in the coronal plasma. Likewise, detected spatial scales of the coronal waves, greater than a few hundred kilometres, exceed typical {plasma kinetic} spatial scales by orders of magnitude. Those estimations justify the description of coronal waves in terms of MHD theory without Hall and electron inertia terms. However, as typical wave scales are comparable to spatial scales of coronal plasma non-uniformities (for example, a typical plasma loop of a coronal active region has the minor and major radii about $10^6$~m and $10^8$~m, respectively), the basic theory of coronal MHD waves is based on accounting for the coronal {inhomogeneity} (e.g., \citep{1983SoPh...88..179E}). 


Essential effects of a field-aligned non-uniformity in temperature and density, resulting in the non-uniformity of the sound, Alfv\'en and fast speeds across the field, are the guided propagation of magnetoacoustic waves along the field because of reflection or refraction, appearance of wave dispersion, and linear transformation of the waves, 
e.g., \citep{MagnetohydrodynamicWaves}. In the low-$\beta$ coronal plasma, parallel phase speeds of magnetoacoustic waves guided by a plasma non-uniformity appear in two intervals associated with slow and fast modes. The slow phase speeds are in the narrow interval between the sound speeds and the so-called tube (or cusp) speed,
\begin{equation} \label{eq:ct}
C_\mathrm{t} = C_\mathrm{Ai}C_\mathrm{si}/(C_\mathrm{Ai}^2 + C_\mathrm{si}^2)^{1/2},
\end{equation}
where $C_\mathrm{Ai}$ and $C_\mathrm{si}$ are the Alfv\'en and sound speeds inside the plasma non-uniformity. Guided slow waves are practically unaffected by the external medium, as far as it is of low $\beta$. Coronal slow waves are often modelled by 1D acoustic equation, i.e., the infinite field approximation. This approach significantly simplifies the theory, but neglects magnetic effects and the perpendicular structure of the wavefront.

In contrast, phase speeds of guided fast waves are determined by the combination of internal and external parameters. In addition, for cylindrical plasma non-uniformities, properties of fast waves are highly sensitive to the azimuthal symmetry of the perturbation, quantified by the azimuthal mode number $m$. One distinguishes between sausage ($m=0$), kink ($|m|= 1$) and fluting($|m| > 1$) modes. In the long-wavelength limit, when the parallel wavelength is much greater than the radius of the cylinder (e.g., the minor radius of a coronal plasma loop), phase speeds of all fast modes except the sausage one approach the kink speed,
\begin{equation}\label{eq:ck}
    C_\mathrm{k} \approx  \Big(\frac{2\zeta}{\zeta+1}\Big)^{1/2} C_\mathrm{Ai},
\end{equation}
with $\zeta = \rho_\mathrm{in}/\rho_\mathrm{ex}$ being the ratio of the internal and external densities, provided $\rho_\mathrm{in} > \rho_\mathrm{ex}$.

Fast waves with parallel wavelengths comparable to the width of the waveguide are subject to strong wave dispersion, as the width is a characteristic scale in the system. Typically, phase speeds decrease with the decrease in the parallel wavelength, while the group speed may decrease or increase  (e.g., \citep{2020SSRv..216..136L}). This dispersion occurs even in the classic low-frequency MHD, without the high-frequency effects such as the Hall or electron inertia effects. The specific dependence of phase and group speeds on the frequency and wave number is determined by perpendicular profiles of the equilibrium plasma parameters, i.e., the density and temperature, and the strength of the magnetic field. Likewise, the geometry of the waveguide, for example, a slab or a cylinder, and the field-aligned current, contribute to the dispersion  (e.g., \citep{2016SSRv..200...75N}). 

In a plasma non-uniformity with a smooth perpendicular profile of the equilibrium Alfv\'en speed, kink waves are subject to linear transformation into local torsional Alfv\'en waves, called \lq\lq resonant absorption\rq\rq\ of magnetoacoustic waves, {see, e.g., \cite{2002ApJ...577..475R, 2006RSPTA.364..433G, 2009A&A...503..213G}}. This effect leads to the damping of kink waves which could be characterised by the exponential damping time 
\begin{equation} \label{eq:td}
\tau_\mathrm{d} \approx \frac{4}{\pi^2} \frac{a}{l} \frac{\zeta+1} {\zeta-1} P_\mathrm{kink},  
\end{equation}
where $l$ is the width of the resonant layer, determined by the steepness of the smooth radial profile of $C_\mathrm{Ai}(r)$ in the vicinity of the location $C_\mathrm{Ai}(r) \approx C_\mathrm{k}$. In the initial phase of the damping, the oscillation envelope may be better described by a Gaussian function, see, e.g., \citep{2013A&A...551A..40P}. 

{The results obtained in terms of the simplified cylindrical model may modify in more complicated geometries, for example, in a bent or sigma-shaped cylinder case or a twisted cylinder. Likewise, in certain circumstances the waves are subject to nonlinear effects.} 
There is a growing number of numerical studies of solar coronal wave processes in terms of 3D full MHD models, see, for example, \cite{2009SSRv..149..153O, 2014RAA....14..805P, 2016A&A...595A..81M, 2019FrASS...6...22P, 2019FrASS...6...38K, 2020ApJ...894L..23M, 2023MNRAS.518L..57L}, {as well as some analytical works, e.g., \cite{2004A&A...424.1065V}}, which can significantly advance the method of MHD seismology. However, one should be cautious, as the outcomes could be affected by various intrinsic numerical artefacts, such as numerical dissipation. 

\section{Wave phenomena observationally detected in the corona}
\label{sec:obs}

In this section we describe several coronal wave phenomena typically used for coronal seismology. Some other phenomena such as propagating kink waves and sausage oscillations are not included in this section because of the space limitations. The interested reader is referred to the recent comprehensive reviews of those topics \cite{2021SSRv..217...76B, 2020SSRv..216..136L}. 

\subsection{Global coronal waves}
\label{sec:eit}

The largest scale MHD wave processes observed in the corona are so-called global coronal waves, often called EUV or EIT waves, or even \lq\lq coronal tsunamis\rq\rq.
They manifest as a single wavefront of an EUV emission intensity disturbance, emanating from an active region almost in all directions along the solar surface (e.g., see \cite{2015LRSP...12....3W} for a comprehensive review). Typical propagation speeds range from 200~\kms to 1,500~\kms. The waves can propagate at large distances, comparable to the solar radius from the epicentre. The lowest propagation speeds are comparable to the sound speed in the coronal plasma, while the highest speeds are comparable to the Alfv\'en speed. 
{Linear and non-linear MHD waves provide one interpretation for these waves. However, some observational features noted in \cite{1999SoPh..190..107D}, such as stationary bright points and coronal dimmings, indicate that there may be a need for alternative, non-wave interpretations. The expulsion of a magnetic plasma may lead to the reconfiguration of magnetic fields and the formation of \lq\lq pseudo waves\rq\rq, misinterpreted as MHD waves. Models include the field line stretching model, e.g., \cite{2005ApJ...622.1202C}, the current shell model, e.g., \cite{2008SoPh..247..123D}, and the reconnection front model, e.g., \cite{2007ApJ...656L.101A}. These models all have different challenges, so several hybrid models with wave and non-wave components have been proposed. Here we focus on the MHD wave interpretation for the global coronal waves.}
In some cases, both the slow and fast wave fronts are detected successively, {see, e.g., \cite{2014SoPh..289.3233L}}. The difference in the propagation speeds indicates that the observed wave motions consist of two types of waves. The rapidly propagating counterpart is associated with the fast magnetoacoustic wave. The propagation of the fast magnetoacoustic component of the global coronal wave is determined by the global 3D profile of the coronal fast magnetoacoustic speed. In a low-$\beta$ plasma, the fast magnetoacoustic speed is about the Alfv\'en speed, and hence its 3D structure is determined by the geometries of the coronal plasma density and the magnetic field strength. In addition, in the low-$\beta$ regime, the fast wave is weakly sensitive to the local direction of the magnetic field. Observations demonstrate that the fast component of the global coronal wave experiences reflection and refraction on coronal non-uniformities of the local Alfv\'en speed, such as active regions and holes. 
The slower component remains subject to intensive debates. In particular, the slower wavefront could be formed due to the successive stretching of closed confining magnetic field lines by an erupting magnetic flux rope in a coronal mass ejection (e.g., \citep{2002ApJ...572L..99C, 2009ScChG..52.1785C}). Generally, drivers of global coronal waves are considered to be coronal mass ejections. 


\subsection{Quasi-periodic fast propagating waves}
\label{sec:qfp}

A more compact version of propagating coronal fast waves are quasi-periodic fast propagating (QFP) waves (e.g., \citep{2022SoPh..297...20S}). Typically, QFP waves appear as a series of arc-shaped wavefronts (or wave trains) of the EUV intensity perturbations, confined to a cone originating at a flaring active region, propagating at speeds exceeding several hundreds \kms, i.e., comparable to the Alfv\'en speed. The waves are detected at heights up to several hundred thousand km, without a significant acceleration or deceleration. Oscillation periods of QFP waves resolved in time and space with modern coronal EUV imagers range from several tens to several hundreds of seconds. Much shorter periods, of about 6~s, have been detected in the white light radiation during a solar eclipse \citep{2003A&A...406..709K}. QFP waves appear as wave trains consisting of a few or several oscillation cycles. Observed properties of QFP waves suggest the existence of two distinct types, narrow and broad QFPs. The main differences of these two types are the angular widths, 10$^{\circ}$--80$^{\circ}$ and 80$^{\circ}$--360$^{\circ}$, and the intensity variation amplitude, less than 10\% and about 10\%--35\%, respectively. Furthermore, narrow QFP waves are seen to propagate along the apparent direction of the magnetic field, while broad waves propagate apparently across the field. A likely reason for the existence of these two types is the difference between the guided and leaky regimes of the fast wave dynamics in a plasma non-uniformity (e.g., \citep{2014A&A...569A..12N}). In this scenario, narrow QFP waves are observed in the waveguiding non-uniformity, while the broad QFP waves are the waves emitted from the waveguide. In the latter case, the plasma non-uniformity acts as a fast magnetoacoustic antenna. 

If the guided and leaky waves are driven by an impulsive energy deposition, i.e., the initial signal is broadband, the dispersion makes different spectral components propagate at different speeds. It leads to the creation of a quasi-periodic wave train \citep{1984ApJ...279..857R}. The modulation of the instantaneous amplitude and oscillation period is determined by the dispersion relation which, in turn, is prescribed by the perpendicular structuring of the plasma. A convenient visualisation of the modulations is provided by the wavelet spectrum. According to theoretical modelling, typical wavelet signatures of QFP waves have characteristic tadpole or boomerang shapes \citep{2021MNRAS.505.3505K}. Signals with these signatures are also detected in quasi-periodic pulsations (QPP) in radio, microwave, white light, and X-ray light curves of solar and stellar flares (e.g., \citep{2021SSRv..217...66Z}), suggesting that QFP waves driven by the flares are responsible for the modulation of the emission. If the driver is periodic, e.g., a sequence of repetitive magnetic reconnection events, i.e., the initial signal is narrowband, the developed fast wave trains are narrowband too. 

\begin{figure}
\centering
\includegraphics[width=0.8\linewidth]{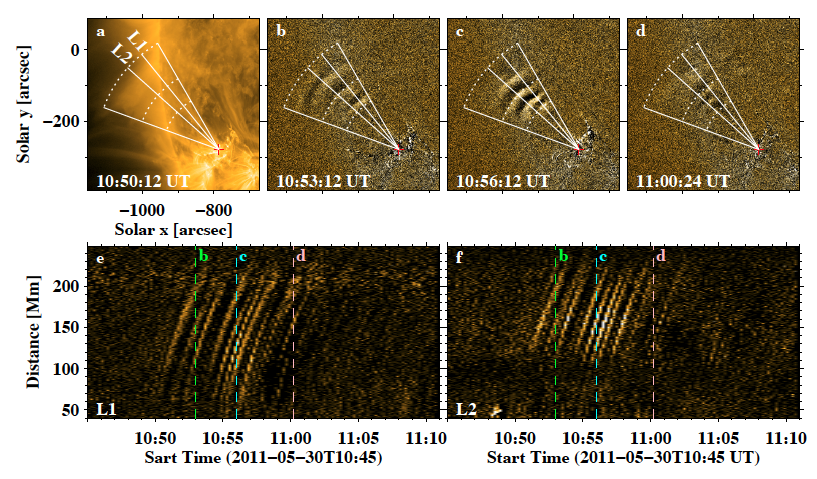}
\caption{QFP waves propagating along a fan-like structure during a C2.8 flare event from 10:48~UT to 11:00~UT on 2011 May 30. Panels (a--d): Original and running difference AIA images in 171\,\AA\ showing the waveguide and snapshots of three consecutive wave trains. The red plus symbol denotes the epicentre of a solar flare. Slits “L1” and “L2” are used to make time--distance maps in panels (e--f) respectively, with a width of 10 pixels. The vertical lines indicate the times of images in panels (b-d). 
\label{fig:qfp}
}
\end{figure}

An example of a QFP wave observed by the instrument SDO/AIA in the 171~\AA\ bandpass in shown in Figure~\ref{fig:qfp}. In this event, there are three wave trains consecutive emanating from the vicinity of a relatively mild solar flare epicentre. In the region of interest, all three wave trains are seen to consist of about four arc-shaped wavefronts. The analysis performed in \citep{2012ApJ...753...53S, 2013A&A...554A.144Y} showed that the projected phase speed of the waves exceeded 800~\kms, which can be estimated by the angle of the diagonal bright lanes in the time--distance map. The dominating oscillation period is about 1~min.  The \lq\lq zigzags\rq\rq\ in the bright lanes are caused by the low temporal resolution.

\subsection{Propagating slow magnetoacoustic waves}
\label{sec:slow}

\begin{figure}
\centering
\includegraphics[width=\linewidth]{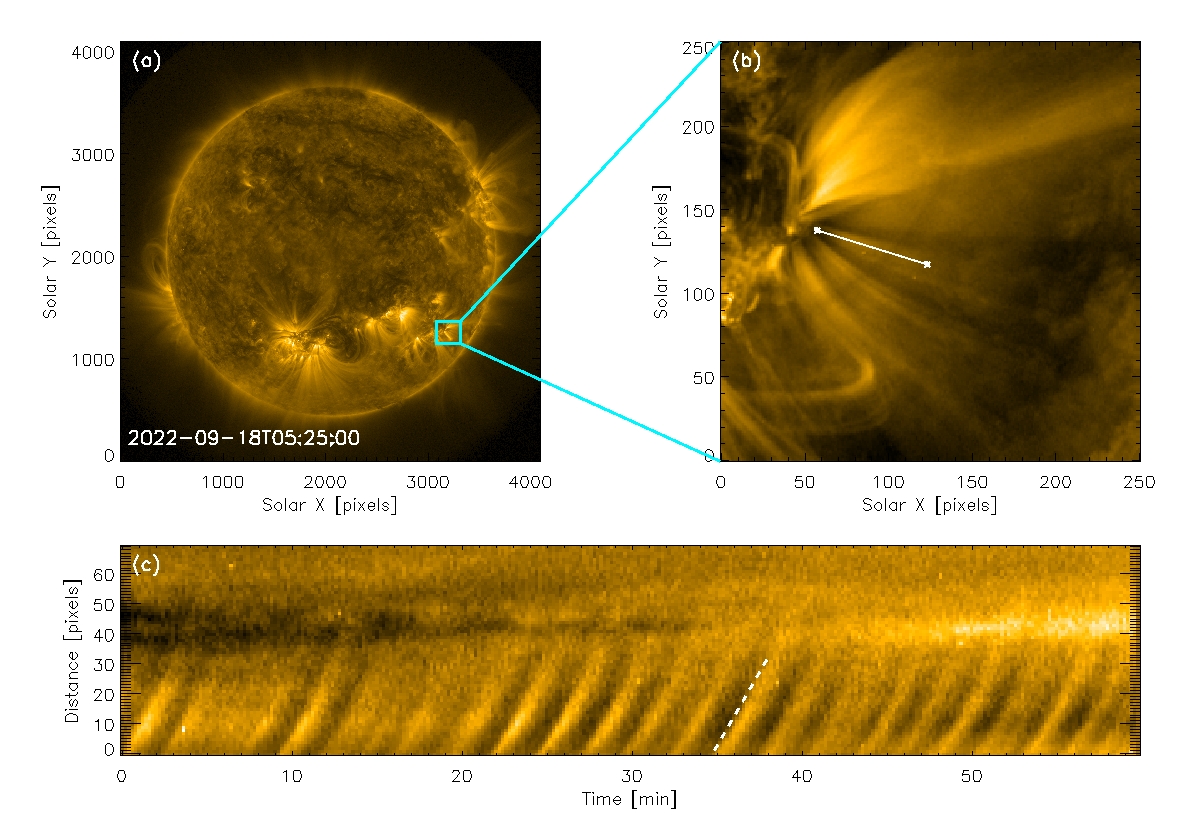}
\caption{Panel (a): The solar disk as seen by SDO/AIA in 171 \AA\, on September 18th 2022 at 05:25 UT. The blue square indicates an active region of interest, with a magnetic fan. Panel (b): A zoomed-in image of the magnetic fan anchored in a sunspot. The white line indicates the slit used to produce the time distance map in panel (c). Panel (c): Time--distance map produced using the slit identified in panel (b) over one hour. The tilted ridges indicate the presence of a propagating slow wave; an example of such a ridge is indicated by the dashed white line.
\label{fig:Slow_Waves}
}
\end{figure}

Another class of longitudinal waves propagating along the apparent direction of the magnetic field are slow magnetoacoustic waves, e.g., \cite{2009SSRv..149...65D, 2021SSRv..217...76B}. The main feature of this class is a low propagation speed which ranges from several tens to a few hundred \kms, i.e., is below or about the sound speed in the corona. The departure from the sound speed is because of the projection of the wave path on the plane-of-the-sky (PoS). Usually, the waves originate from the chromosphere, and are detected in the lower part of coronal loops or polar plumes. Very often, waves of this class are detected in magnetic fans spreading out of sunspot umbrae and pores. Typical oscillation periods are from a few to several minutes. Normally, shorter periods, about 2--5~min, are observed in active regions, while plumes guide slow waves with periods longer than 6--7~min, in some cases, up to 20~min.  
The waves may be generated for several tens of oscillations cycles with almost constant period, resembling a maser, which also makes slow waves different from QFP waves. Another difference with QFP waves is the short damping distance. The waves are usually detected at low heights: the slow waves disappear from observations within 15--20 thousand km. 

Figure~\ref{fig:Slow_Waves} presents a typical example of propagating slow waves in a coronal magnetic fan. The time--distance map (see Section~\ref{sec:tdm}) shows a propagating slow wave pattern which is clearly resolved up to the distance about 35 pixels from the footpoint. Taking that the pixel size of the instrument is about 440~km, the projected detection length is about 15 thousand km. In the time domain the waves show a high degree of coherency. The propagating speed projected on the PoS can be estimated by the tilting angle of the diagonal ridges in the map. In that example, the phase speed seems to increase slightly with height, which could be attributed to the change of the angle between the wave path, i.e., the local magnetic field, and LoS. For the diagonal dash line, we determine that the wave disturbance travels about 35 pixels in about 3 min. It gives us a propagation speed of about 85~\kms. 

In some cases, QFP waves and slow waves are detected to propagate simultaneously along the same coronal plasma structure, e.g., \cite{2015A&A...581A..78Z}. The slow waves propagating at the projected speed of about 55--105~\kms\ are persistently present at lower heights, up to about 20,000~km from the origin, while the fast waves appear as two distinct wave trains with speeds about 800--900\kms, reaching distances exceeding 150,000~km.

\subsection{Standing and sloshing slow waves}
\label{sec:sumer}

Another manifestation of coronal slow magnetoacoustic waves is the bouncing variations of the EUV or thermal microwave emission intensity along coronal loops. If LoS is oblique to the plane of the loop, these wave motions appear also as oscillatory Doppler shifts of coronal emission lines, registered near the top of the loop, see \citep{2011SSRv..158..397W, 2021SSRv..217...34W} for a recent comprehensive review. In some cases, the {Doppler} velocity and intensity oscillations are detected simultaneously, and show a quarter-period phase shift, indicating the standing nature. The oscillation periods are typically longer than several minutes, exceeding in some cases 15 minutes. The ratios of the wavelength estimated as double the length of the oscillating loop to the oscillation period give the phase speed of several hundreds \kms, up to about 500~\kms. Those values correspond to the sound speed in the loop, estimated by the plasma temperature. This property indicates the slow magnetoacoustic nature of this wave phenomenon.  
The rareness of the simultaneous detection of the oscillations in Doppler shift and intensity is linked with the spatial separation of the locations of the maximum amplitudes in a standing slow wave. In the fundamental slow mode, the parallel velocity oscillations have a maximum at the top of the loop, while the density oscillations have there a node. 

In the vast majority of cases, oscillations of this kind have been detected in loops filled in with hot plasma, with the temperature of $6.3\times 10^6$--$14 \times 10^6$~K. There is a tendency, that in cooler loops {the periods are longer compared to hotter loops} \citep{2019ApJ...874L...1N}. The oscillations are subject to rapid damping, with the exponential damping time being about the oscillation period. {Statistically, the damping time scales almost linearly with the oscillation period \citep{2019ApJ...874L...1N}}. The initial relative amplitude of the oscillations could exceed a few tens of \%. Oscillations with similar properties have been detected in the white-light emission of stellar flares \cite{2016ApJ...830..110C}.    

A wave phenomenon similar to standing slow waves is an EUV intensity perturbation bouncing between the footpoints of a loop.
In the time--distance map (see Section~\ref{sec:tdm}) made for a curvilinear slit directed along the loop, these wave motions exhibit a \lq\lq zigzag\rq\rq\ pattern, which is clearly distinct from a chessboard pattern of a standing wave. This kind of a slow wave motion is called a sloshing oscillation \cite{2013ApJ...779L...7K, 2016ApJ...826L..20R}.  The oscillation periods and decay times of sloshing oscillations are similar to those of standing slow oscillations. The apparent phase speed of the bouncing bright spot matches the sound speed in the loop. {The exact formation mechanism of sloshing oscillations is not yet known. This phenomenon is distinct from a reflected propagating slow wave in the loop, as sloshing oscillations do not manifest the frequency-dependent damping, with all spatial harmonics decaying at approximately the same rate \citep{2019ApJ...874L...1N}. On the other hand, in some studies sloshing oscillations were shown to convert into a typical slow standing wave pattern after several footpoint reflections, regulated by the competition between thermal conduction and compressive viscosity in wave damping, with standard or anomalous transport coefficients \citep{2018ApJ...860..107W, 2021ApJ...914...81K}.}

\subsection{Kink oscillations}
\label{sec:decay}

Perhaps one of the most intensively studied oscillatory phenomena in the solar corona is kink oscillations of coronal plasma loops (e.g., \citep{2021SSRv..217...73N}). The oscillations occur as standing oscillatory transverse displacements of the loops from an equilibrium, or Doppler shift oscillations of coronal emission lines along LoS. Kink oscillations appear in two different regimes: large-amplitude rapidly decaying oscillations and low-amplitude decayless oscillations. In both regimes, the oscillation periods are typically several minutes, reaching about 30~min in longer loops. Transverse displacements of different segments of the loop are either in phase with each other or in anti-phase. The oscillation periods scale linearly with the length of the oscillating loop. These two properties clearly indicate the standing nature of the oscillations. The wavelength is determined by double the length of the loop, divided by the parallel harmonic number of the oscillation. The lowest, or first (or fundamental), parallel harmonic has the longest oscillation period. Higher parallel harmonics are detected even rarer than the fundamental harmonics. In the decaying regime, typical initial displacement amplitudes projected on the PoS are several thousands~km, while in the decayless regime, the stationary amplitudes are much lower, less than a few hundred~km. 

In the decaying regime, the oscillation lasts for several oscillation cycles only. Usually, the amplitude shows a steady decay in time, which could be approximated by an exponential or another decreasing function with a certain characteristic damping time (see Fig.~\ref{fig:deck2} and cf. Equation~\ref{eq:td}). A vast majority of decaying kink oscillations are excited by a displacement of the loop from an equilibrium by a plasma eruption occurring nearby \cite{2015A&A...577A...4Z}.  Decaying oscillations are rather rare, with only several hundred events detected during a full solar cycle \citep{2019ApJS..241...31N}.

Decayless oscillations last for up to several tens of oscillation cycles, and their amplitude does not show any systematic evolution. Decayless kink oscillations seem to be ubiquitous. {The instantaneous amplitude shows some variations in time, and in some time intervals the oscillation amplitude is seen to grow \citep{2012ApJ...751L..27W}.} Sometimes, decaying kink oscillations are not seen to damp to zero, but instead approach the amplitude of the decayless oscillation \cite{2013A&A...552A..57N}. This behaviour resembles a dynamic system characterised by a phase portrait with a limit cycle, i.e., a self-oscillator \citep{2016A&A...591L...5N, 2020ApJ...897L..35K}. In this scenario, the oscillation period is determined by the natural oscillation period determined, in the case of kink oscillations, by the ratio of the wavelength and the kink speed (\ref{eq:ck}), while the oscillation damping is counteracted by some external energy supply. The energy which sustains the oscillation can be supplied by quasi-static plasma flows in the external medium, similar to the movement of a bow across a violin string. In the solar atmosphere, those motions could occur near the loop's footpoints, for example, the low-frequency part of the red-spectrum granulation and super-granulation flows, and Evershed flows, or be present in the corona. In another scenario, decayless oscillations are sustained by perpetual random movements of the footpoints, caused by, e.g., the granulation. This latter mechanism has a shortcoming, as it should produce randomly polarised oscillations, which contradicts recent quasi-stereoscopic observations \citep{2023NatCo..14.5298Z}.  

\section{How to obtain, process and analyse observational data}
\label{sec:data}

As a typical example of the acquisition, processing and analysis of observational data performed in MHD seismology, consider analysis of EUV imaging data obtained with the Atmospheric Imaging Assembly (AIA) instrument \cite{2012SoPh..275...17L} on the Solar Dynamics Observatory spacecraft. Below we describe several typical steps which are carried out first online and then in the Interactive Data Language (IDL\footnote{\url{https://www.nv5geospatialsoftware.com/Products/IDL}}) environment with the use of the Solar Software (SSW) package \citep{1998SoPh..182..497F}. AIA data is in the open access. Python users can use the SunPy ecosystem \citep{2023FrASS..1076726B}. This example gives the reader an idea of the acquisition, processing and analysis of data sets obtained with other imaging instruments, such as the extreme ultraviolet imagers STEREO/EUVI \footnote{Tutorials for EUVI data processing and analysis are available at \url{https://secchi.nrl.navy.mil/data-analysis}} and SolO/EUI \footnote{User manual of EUI data is available at \url{https://www.sidc.be/EUI/data/releases/202301_release_6.0/release_notes.html}}, \cite{2008SSRv..136...67H, 2020A&A...642A...8R}, respectively. 

\subsection{Data acquisition and pre-processing }
\label{sec:aia}
 For the identification of the event of interest, it is convenient to use the {JHelioviewer \citep{2017A&A...606A..10M}} environment (\url{https://www.jhelioviewer.org/}, this website has a detailed tutorial). The initial selection of the region and time interval of interest can be based on known locations and times of some dynamic events, such as flares and various eruptions, or, contrarily, during the quiet periods of solar activity. Similarly, regions around sunspots, pores, faculae, prominences, coronal holes, active regions, etc., where oscillatory processes of interest are expected, could be chosen.  

\begin{figure}
\centering
\includegraphics[width=0.35\linewidth]{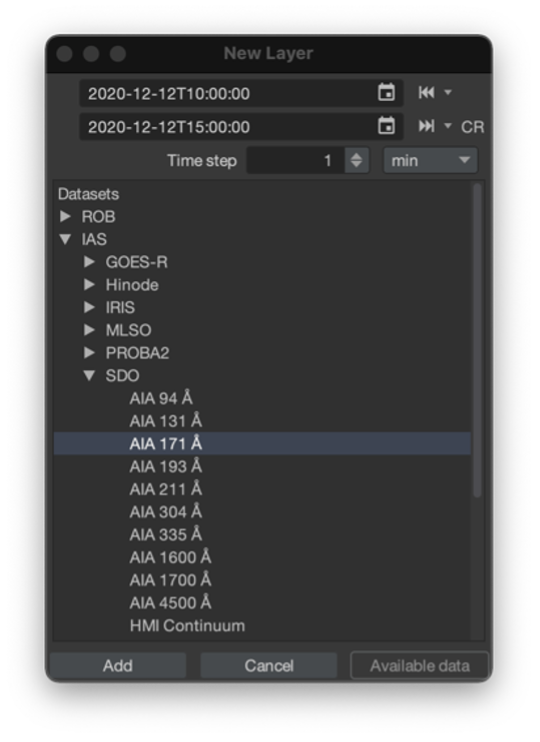}
\includegraphics[width=1.0\linewidth]{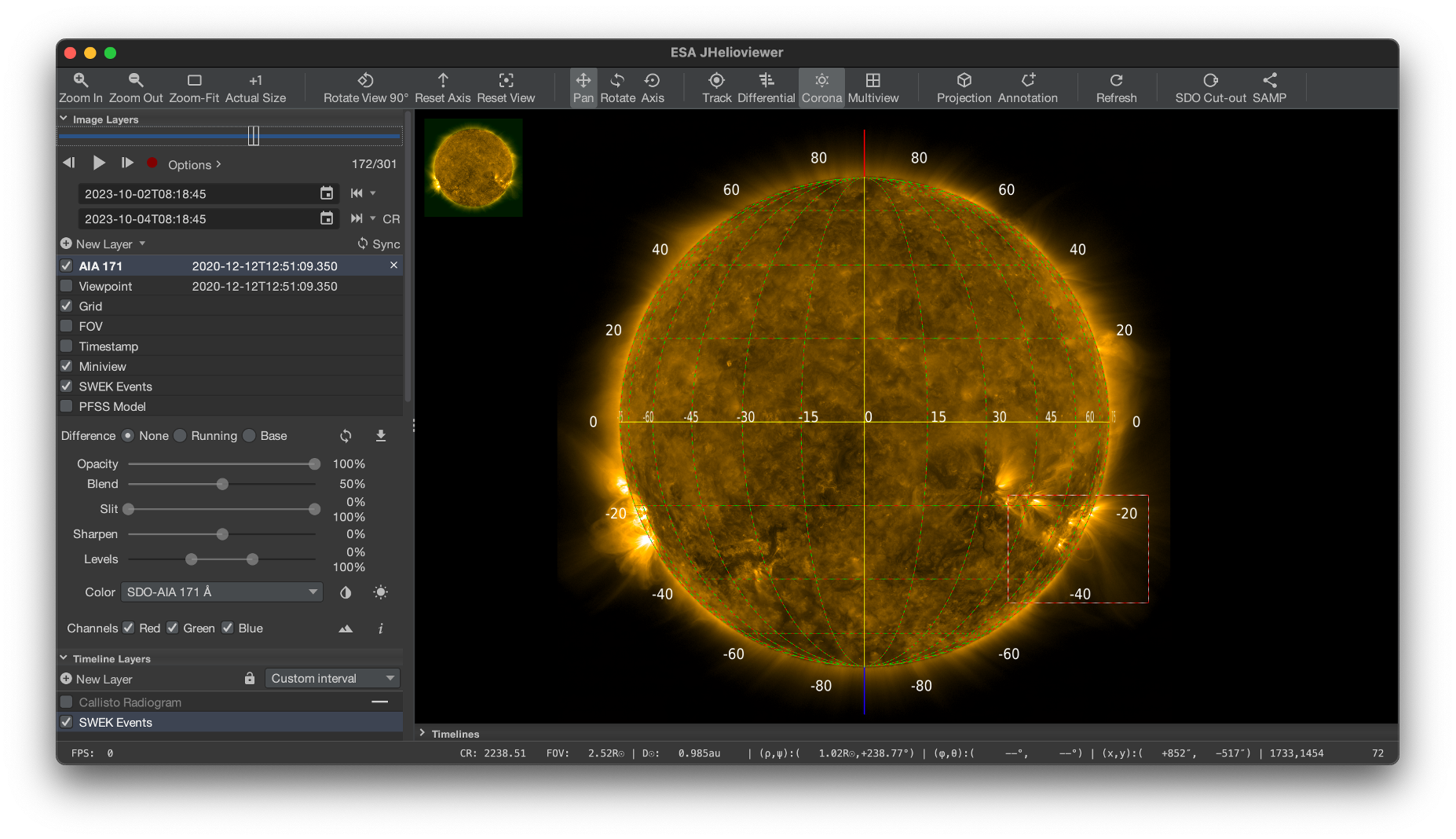}
\caption{{JHelioviewer} interface to browse data. {Up}: tab of \lq\lq New Layer\rq\rq\ showing the parameters of the region and time interval of interest. {Bottom}: The full-disk image at the selected time, with the red rectangle showing the region of interest
\label{fig:jhv}
}
\end{figure}

The first step is to click \lq\lq New Layer\rq\rq\ or press \lq\lq Control + N\rq\rq, select \lq\lq SDO\rq\rq\ then \lq\lq AIA\rq\rq, set the observational wavelength, for example, 171~\AA, which gives high contrast coronal images, the time step, e.g., 12~s, and a rough time range (see Figure~\ref{fig:jhv}). One can select images of different wavelengths simultaneously. This step results in selecting a data cube, i.e., a 3D array with two spatial dimensions in the PoS (i.e., frames) and one time dimension. The data cube could be a stack of either original images, or their differences. The image difference signals can be either running, with each frame being the difference between the current image and the previous one, or base, when each frame is a difference of the current image and some base one which is the same for all the frames. The choice of the difference is made by selecting \lq\lq Running\rq\rq\ or \lq\lq Base\rq\rq\ in the toolbox shown on the right. The running difference and base difference data cubes could be considered as an effective time derivative and detrending of the original data cube, respectively. 

It is worth trying to identify oscillatory motions in the selected data set visually, running the movies constructed by the data cubes. It will ensure that all features of interest, i.e., all wave paths or all oscillating structures are in the field-of-view of the selected data cube. For example, the data cube with decayless kink oscillations of a coronal loop on 2020-12-12 starts at 10:30:00~UT and ends at 13:50~UT, the centre of the region of interest (RoI) is [+838, -477] in arcsec, the width and height of RoI are both 500 pixels, and the time cadence is 12\,s.

\begin{figure}
\centering
\includegraphics[width=1.0\linewidth]{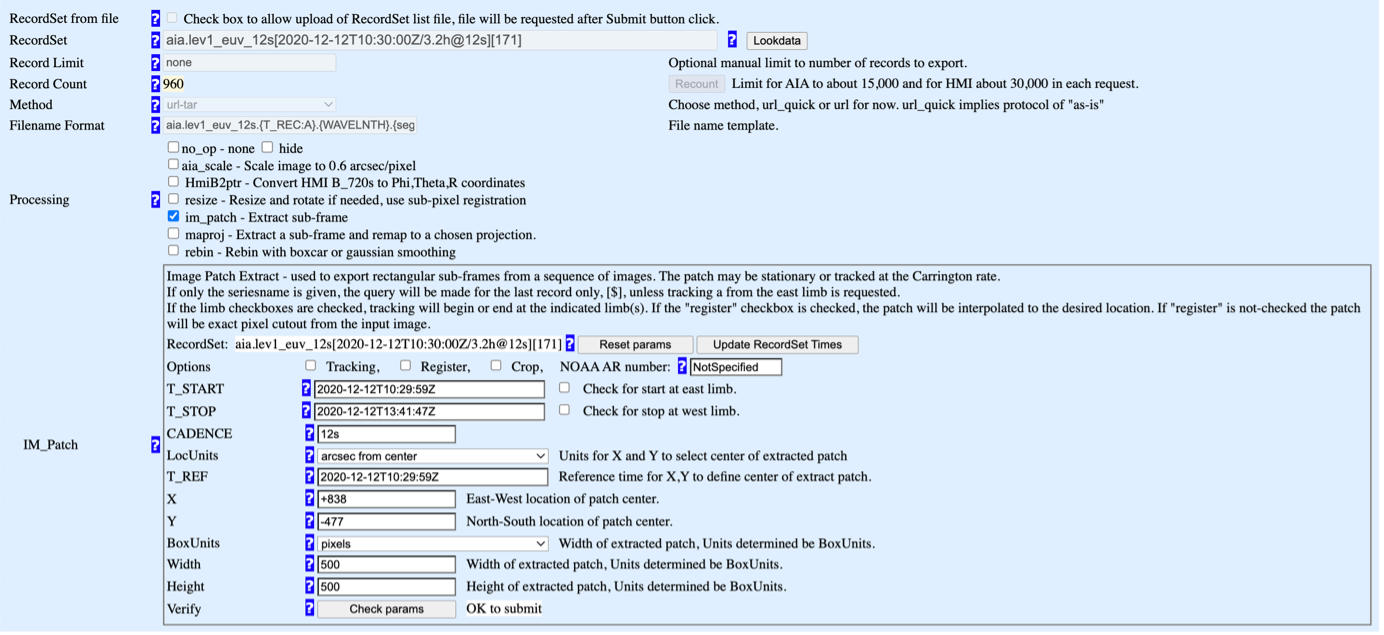}
\caption{A screenshot of the JSOC export web page, showing the \lq\lq Enable Processing—im\_patch\rq\rq\ form to extract a region-of-interest for further analysis.
\label{fig:jsoc}
}
\end{figure}

It is recommended to download AIA data from the Joint Science Operation Centre (JSOC) centre. On the JSOC Lookdata website (\url{http://jsoc.stanford.edu/ajax/lookdata.html}), in the tab of \lq\lq Series Select\rq\rq, click \lq\lq fetch the observational series name list\rq\rq\ and then click \lq\lq aia\_lev1\_euv\_12s\rq\rq. After jumping to the tab \lq\lq RecordSet Select\rq\rq\ (see Figure~\ref{fig:jsoc}), insert the selected time, duration, and wavelength, e.g., \lq\lq [2020-12-12T10:30:00Z/3.2h@12s][171]\rq\rq\ for the example mentioned above. To obtain data at multiple wavelengths, which correspond to the emission from a plasma with different temperatures that can be used for differential emission measure analysis (see Section~\ref{sec:dem}), add the additional wavelengths in the last bracket, e.g. \lq\lq [171,193,211]\rq\rq. Then click the \lq\lq Export Data\rq\rq\ tab, and then \lq\lq export\rq\rq, which will direct to the export page.

In the export page, select \lq\lq url-tar\rq\rq\ in \lq\lq Method\rq\rq\ and \lq\lq FITS\rq\rq\ in \lq\lq Protocol\rq\rq\ to compress the data in the Flexible Image Transport System (FITS) format into a tape archive (tar) format file, and deliver with URL link. To download full-disk images, input your email address and submit the request. New users should register with an email address. To register, put your email address in \lq\lq Notify\rq\rq, wait for the activation email and follow the instructions therein. To extract cutouts, i.e., partial RoIs, click \lq\lq Enable Processing\rq\rq, then click \lq\lq im\_patch\rq\rq\ and input the centre, height, and width of the RoI. In the study of oscillatory processes discussed in this tutorial, it is advised to {either unselect \lq\lq tracking\rq\rq\ or select both \lq\lq tracking \rq\rq\ and \lq\lq register\rq\rq} to avoid the 1-pixel sawtooth artefacts generated by JSOC’s automatic de-rotation. However, this operation is not necessary if the RoI is near or off the limb. Then click \lq\lq check parameters\rq\rq\ and \lq\lq submit\rq\rq. When JSOC has processed the query, the user receives an email with the link to the prepared data ready for downloading, compressed in a tar file. 

Sub-frames can also be extracted using Lockheed Martin Solar and Astrophysics Laboratory cutout service (\url{https://www.lmsal.com/get_aia_data/}), or the Virtual Solar Observatory (\url{https://sdac.virtualsolar.org/}), or with the use of the \verb|ssw_cutout_service.pro| function in IDL {or sunpy.net.Fido package in Python}. More details are available in Section~5.4 of the Guide to SDO Data analysis by M.~DeRosa \& G.~Slater (\url{https://www.lmsal.com/sdodocs/doc/dcur/SDOD0060.zip}), or P.~Young’s SDO analysis guide (\url{https://pyoung.org/quick_guides/pry_sdo_guide.html}).

\subsection{Data processing }
\label{sec:proc}

The AIA data downloaded from the JSOC centre is Level 1, which has been flat-fielded and processed to remove bad pixels and spikes. 
The Level 1 data should be processed with the \verb|aia_prep.pro| routine in IDL to obtain the Level 1.5 format. This procedure co-aligns the images, seeking the currently best-known pointing information of the instrument; accounts for the roll angle; normalises the image by the exposure time; and re-scales them to a common plate scale. 
In Python, a similar procedure can be performed with \verb|aiapy| in the SunPy package, see \url{https://aiapy.readthedocs.io/en/stable/preparing_data.html}.
The data obtained from the LMSAL cutout service are in level 1.5, so there is no need for \verb|aia_prep|. 

The next important step is the removal of the solar rotation. The Sun rotates from east to west, hence a target of interest moves in the PoS at a non-steady speed which is determined by the local heliographic coordinates. Thus, there may be a need to account for the effect of the solar rotation, keeping the initial location of the target in the RoI. This operation must not affect the target's own movements, such as transverse oscillations, in the RoI. Usually, the de-rotation which removes the projected motions caused by the solar rotation is required for targets on the solar disk near the central meridian, while is not needed for off-limb plasma features. This operation is based on re-sampling the images with an updated coordinate system with respect to the reference frame. In IDL, the \verb|drot_map.pro| routine is used. There are also other tracking methods such as \verb|sdo_track_object.pro| and  \verb|sdo_prep.pro|. {Note that the de-rotation should not be done if a user has selected ``tracking" and ``register" options while exporting data from JSOC.}

In Python, \verb|sunpy.map.reprojected_to| and \verb|propagate_with_solar_surface| in \verb|sunpy.coordinates| is used with the reference date and map centre, to rotate a map differentially. Alternatively, \verb|RotatedSunFrame| in \verb|sunpy.coordinates| can be used to de-rotate the coordinate for the image. Details are available in the SunPy online tutorial: \url{https://docs.sunpy.org/en/stable/generated/gallery/differential_rotation/index.html}.

\subsection{Making time--distance maps}
\label{sec:tdm}

\begin{figure}
\centering
\includegraphics[width=1.0\linewidth]{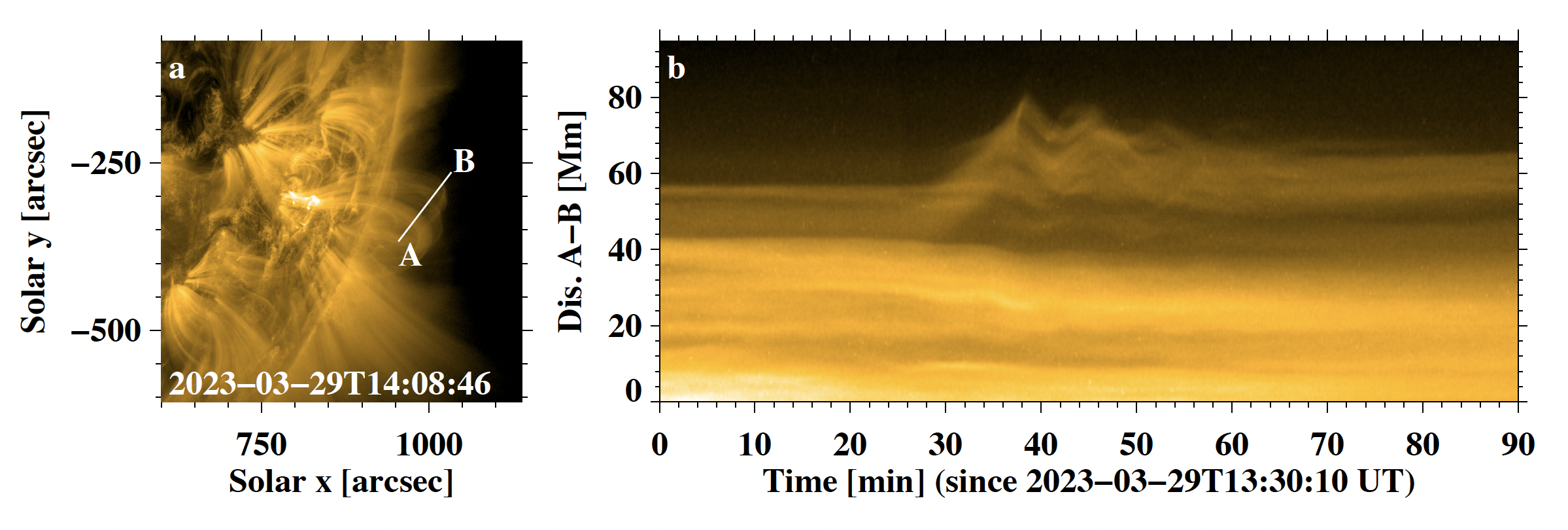}
\caption{An example of a time--distance map, showing an oscillatory pattern representing a kink oscillation of a coronal plasma loop after an M-class flare on 2023 March 29th from 13:30 to 15:00~UT. Panel a: an AIA 171\,\AA\ RoI including the oscillating loop. The white line indicates the slit used for making a TD map displayed in panel b.
\label{fig:tdm}
}
\end{figure}

A common approach in detecting and analysing coronal oscillatory processes is with the use of time--distance (TD) maps. It allows us to reduce the investigation of a 3D data cube to the analysis of a 2D array. The TD map is a 2D array with one time coordinate, and one spatial coordinate along a chosen slit, showing the time variation of the emission intensity along the slit with the time. Making a TD plot is a two-step procedure: firstly, one needs to determine spatial coordinates of the slit in the spatial domain; and, secondly, take the intensity values along the slit. Thus, we obtain a rectangular bar with the length and width of the slit. The width is at least one pixel wide, or more. Then the bars are stacked along their long sides, making the map. For kink oscillations, this procedure is illustrated in Figure~\ref{fig:tdm}.
Figure~\ref{fig:Slow_Waves} demonstrates a time--distance map which reveals a persistent and almost coherent slow wave. 

The position of a slit can be defined in a RoI after the visual detection of an oscillation or its anticipation. Slits can be chosen to be straight or curved, and their orientation depends upon the oscillatory process of interest. For longitudinal waves, such as propagating and standing slow magnetoacoustic waves (see Sections~\ref{sec:eit}---\ref{sec:sumer}), the slit should be directed along the wave path. For transverse waves, such as kink waves (see Section~\ref{sec:decay}), the slit should be positioned across the oscillating structure, in the direction of its transverse displacements. Often, a slit is designed to have a width of several pixels, averaging the intensity over the width, to increase the signal-to-noise ratio. 

If the slit is correctly chosen, the TD map demonstrates the presence of standing and propagating (including sloshing) wave motions in the RoI, as variations of the intensity in the map image. As the next step, a skeleton of the pattern highlighting the wave motion should be determined. This can be done by determining the coordinate of the maximum intensity across the lane of the pattern, in each time frame. In the case of a kink oscillation, the location of the maximum intensity will represent the centre of the oscillating loop. Alternatively, a location of the maximum intensity gradient could be determined. 

The identified path can be fitted with a certain analytical expression either suggested by theory, or guessed. For example, in the case of decaying kink oscillations it could be a harmonic function decaying exponentially according to Eq.~(\ref{eq:td}), see Figure~\ref{fig:deck2}. Sometimes, the fitting procedure begins with removing an aperiodic trend, which could be determined either as a low-order polynomial function, or simply by smoothing the signal, e.g., with the \verb|smooth| function in IDL {or with uniform\_filter1d function from scipy.ndimage package in Python.}

\subsection{Analysing sub-resolution signals in imaging data}
\label{sec:mm}
Imaging telescopes used for the observational detection of solar coronal wave processes usually have high spatial and temporal resolution. For example, SDO/AIA has the pixel size of 0.6\,arcsec, corresponding to the linear scale of about 440~km on the solar surface, and a cadence time of 12~s in the EUV channels. However, coronal oscillation amplitudes are often near the resolution threshold. For example, typical displacement amplitudes of decayless kink oscillations are about 0.3\,arcsec or smaller. Likewise, sub-resolution amplitudes are often registered in high spatial harmonics of decaying kink oscillations, and at the loop segments near the nodes. Typical oscillation periods of QFP waves are shorter than one minute. Thus, there is a need for data processing techniques which allow one to study low-amplitude oscillatory processes. 

The Motion Magnification (MM) technique, based on Two-dimensional Dual Tree Complex Wavelet Transform, DTCWT, \cite{2005ISPM...22..123S}, amplifies transverse quasi-steady low-amplitude motions in the PoS. The amplification coefficient is limited by the instrumental noise and jitter, and can, for example, reach 6 for SDO/AIA data. The MM technique allows for amplifying oscillatory motions with a broad range of periods, typically of one or two orders of magnitude. In the time--distance maps processed with MM, oscillatory displacements are readily detectable by eye.  

The principle of DTCWT-based MM is described as follows. The 2D DTCWT is a wavelet spatial transform. It decomposes an input image sequence into a set of complex-valued high-pass images (6 orientations) of different scales and low-pass residuals. The phase difference of high-pass images is linearly proportional to the object displacements. The relative variation of phase is computed by subtracting the phase trend obtained by smoothing the time series of phase with a flat-top window of time width $w$. This step excludes the transverse motions with timescales longer than $w$. Then, the detrended phase is amplified by a magnification factor k through multiplying. Next, the magnified images are reconstructed via inverse DTCWT using the modified phase, which is the sum of the phase trend and magnified relative phase. The intensity and periods of an object are invariant before and after MM.

This method involves two user-defined keywords: magnification factor $k$ and smoothing width $w$. The former determines how many times the amplified amplitude is bigger than the original. The latter decides the bandwidths of motions to be amplified. Being tested on both synthetic data and real solar imaging data \citep{2016SoPh..291.3251A, 2021SoPh..296..135Z, 2022SSRv..218....9A}, this technique is found that\\ 
(a) the linear scaling of magnified and original amplitude works well in the low-amplitude range from 0.01--1 pixels;\\ 
(b) the smoothing width w should be longer than the estimated motion period;\\ 
(c) artificial cross-influence could be generated where neighbouring structures that are too close to each other become overlapped after MM;\\ 
(d) the effect of background noise or other irregular movements on the MM performance is minor.

The realisation of the 2D DTCWT-based MM is available in \url{https://github.com/Sergey-Anfinogentov/motion_magnification}. 
It can be {called from IDL or Python code}. An example usage is demonstrated in the given link. Prior to the installation of MM, you should install Python 3 and DTCWT for Python (\url{https://github.com/rjw57/dtcwt}). An IDL user should download the MM package and then compile the procedure 
\verb|magnify_2d.pro| 
to make it ready for use. For Python users, download the package and \verb|magnify.py| is the function provided, which can be called directly under its directory.

Although MM is designed for solar imaging data, it can also be used in image sequences, i.e., movies generated by a digital video camera. The best situation is that the boundary of the object which performs oscillatory movements has high contrast. The following is the example usage.

First, stack the image sequence or convert the movie into a 3D data cube (with two spatial dimensions representing the PoS, and the 3rd dimension being the time) as input data for MM. The size of the spatial dimensions of images must be even, in the units of pixels. The larger the size, the longer the processing time. Note that this also will be limited by the computer memory. The length of the sequence could be as long as two thousand frames.  



Then, determine the magnification factor $k$, depending on the original and target amplitude of the transverse oscillatory displacements in the PoS. For displacement of 0.1--1 pixels, the value of $k$ usually ranges from 1--10. Typically, we select $k=5$ for the first attempt. Next is the value of smoothing width $w$ in the unit of time frame, which should be longer than the expected oscillation period of the motion. For example, if the anticipated periodicity is roughly 5~minutes, that is 25~frames, provided the time cadence is 12~seconds, then $w = 35$. 

Last, call the procedure in IDL or Python with the data and the coefficients as input. The procedure returns a \lq\lq magnified\rq\rq\ data cube which has the same dimension as the input data.

In IDL:
\begin{verbatim}
IDL> .compile -v '/your_path_name/motion_magnification/magnify_2d.pro'
% Compiled module: RANDOMSTRING.
% Compiled module: LOAD_CUBE.
% Compiled module: SAVE_CUBE.
% Compiled module: MAGNIFY_2D.     
IDL> magnified_data = magnify_2d(data, k, w)
\end{verbatim}
In Python:
\begin{verbatim}
from magnify import *
magnified_data = magnify_motions_2d(data, k, w)
\end{verbatim}
\noindent where \verb|data| is the 3D input original data, and the resulting magnified data is \verb|magnified_data|.

\begin{figure}
    \centering
    \includegraphics[width=0.95\columnwidth]{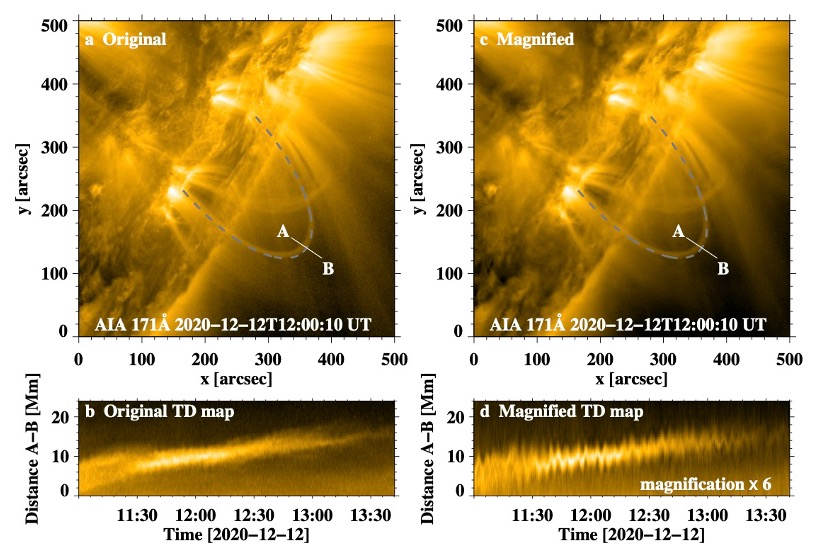}
    \caption{Example application of the motion magnification to an AIA data cube. Left: the original image and time--distance map made using the slit A--B. Right: magnified image and TD map. The magnification factor is 6. The size of the data is $600 \times 600 \times 811$.
    \label{fig:mm}
    }
\end{figure}

An example result is shown in Figure~\ref{fig:mm}. After MM, the image quality is degraded slightly.

\subsection{Bayesian best-fitting of empirical data}
\label{sec:mcmc}

As MHD seismology is based upon the interplay of empirical data and theoretical modelling, an important element of this method is the comparison of theoretically predicted and observationally determined functional dependencies of various physical parameters. Often, the theoretical dependencies (or \lq\lq models\rq\rq) have a number of free parameters, which complicates their determination by standard techniques for the approximation, such as the least-square methods. This difficulty could be mitigated by the Bayesian approach implemented through the Monte Carlo Markov chain sampling. A detailed justification of the use of a Bayesian probabilistic approach to MHD seismology of coronal plasmas, and a comprehensive review of recent results are given in \citep{2022FrASS...926947A, 2022SSRv..218....9A}. Here, we discuss a practical tool, the Solar Bayesian Analysis Toolkit (SoBAT) software package \citep{2021ApJS..252...11A}, available in \url{https://github.com/Sergey-Anfinogentov/SoBAT}.

SoBAT is designed specifically for the Bayesian analysis of solar observational data, in general, and tasks of coronal seismology, in particular. In addition to the reliable estimation of user-supplied model parameters and their credible intervals, it allows for a quantitative comparison of different competing models by means of the Bayesian factor, $K_{i,j}$. The latter is determined by the ratio of the Bayesian evidence of model \lq\lq i\rq\rq, $B_i$, to the Bayesian evidence of model \lq\lq j\rq\rq, $B_j$, as $K_{i,j} = 2\ln{B_i/B_j}$, and higher values of the Bayesian factor $K_{i,j}$ indicate the preference of model \lq\lq i\rq\rq over model \lq\lq j\rq\rq. In this section, we overview practical steps that one needs to make to fit the time series of interest with an \textit{a priori} prescribed multi-parametric theoretical model and for model comparison with SoBAT v.0.3.1.

\begin{figure}
    \centering
    \includegraphics[width=0.95\columnwidth]{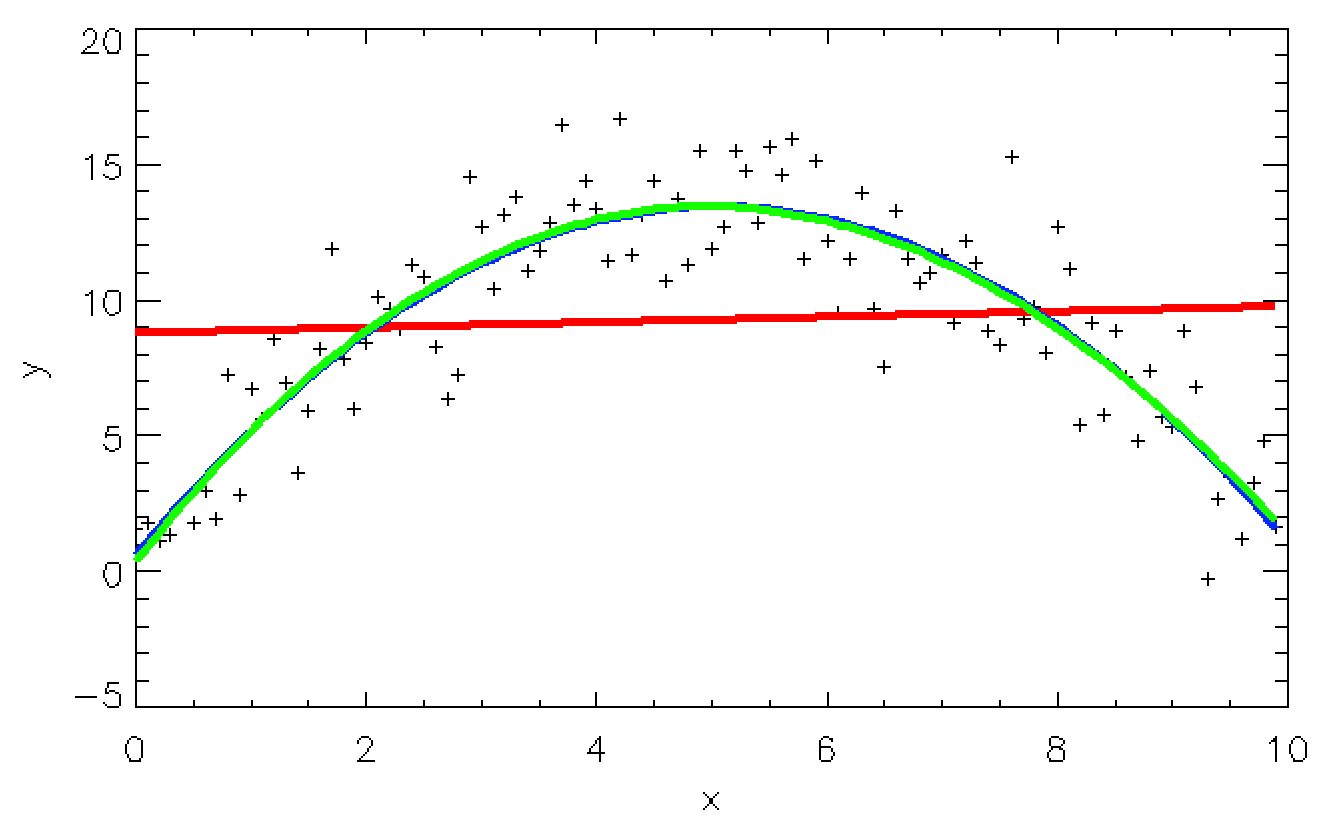}
    \caption{Example of the application of SoBAT to a synthetic parabolic signal with random noise (black crosses). The best-fitting curves obtained with the linear, parabolic, and cubic models are shown in red, blue, and green, respectively. 
    \label{fig:mcmc_synth}
    }
\end{figure}

\begin{figure}
\centering
\includegraphics[width=0.7\linewidth]{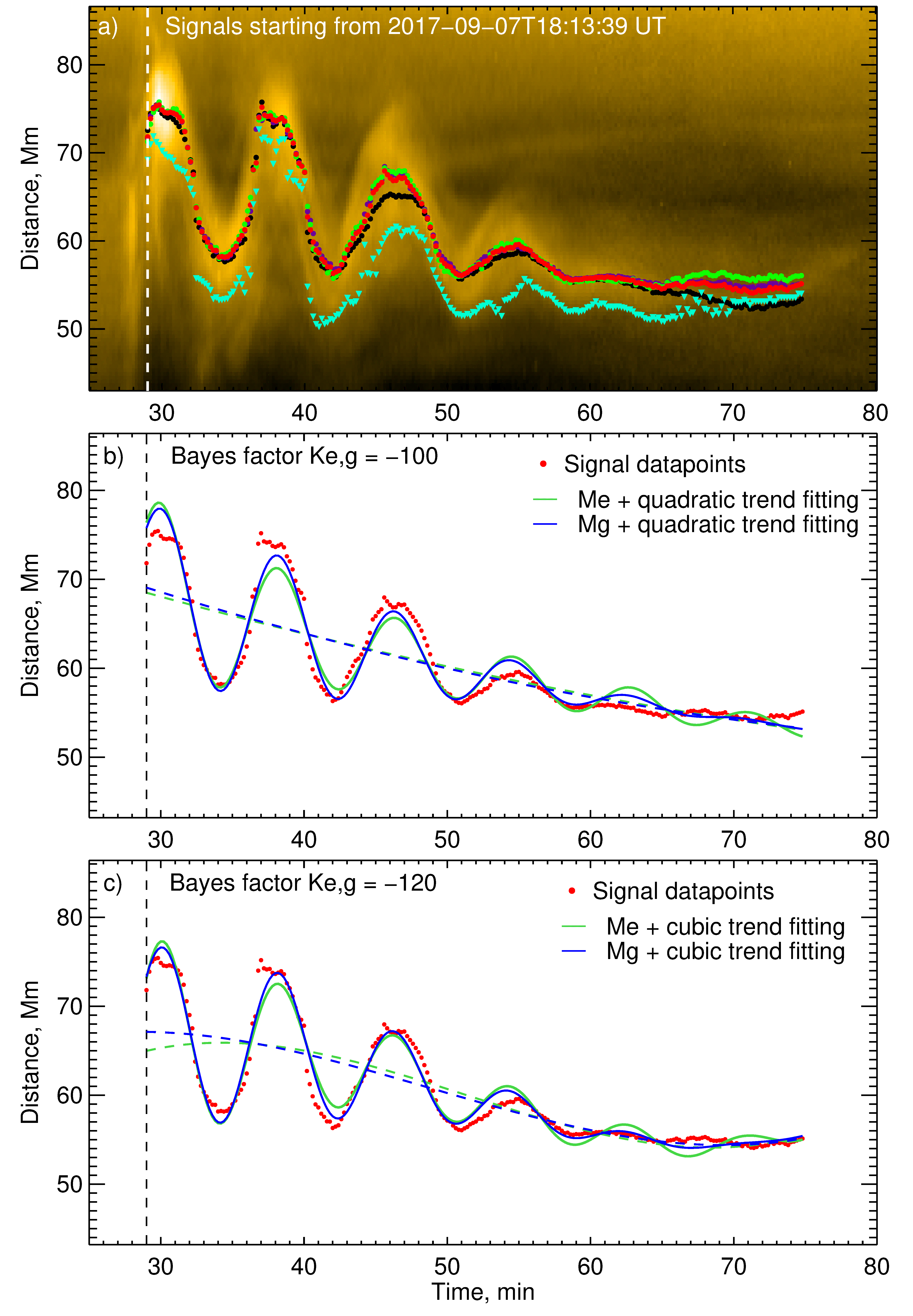}
\caption{{Time--distance map of the kink-oscillating loop highlighted in Fig.~\ref{fig:cor} (panel a). The symbols of different colours represent the loop displacements tracking with five different algorithms described in \citep{2023MNRAS.525.5033Z} and Sec.~\ref{sec:proc} .} Best-fitting the identified kink-oscillating loop displacement with the SoBAT software package, by a harmonic function with an exponential decay ($M_e$) and Gaussian-exponential decay ($M_g$), and quadratic and cubic background trends (panels b and c, see Sec.~\ref{sec:mcmc}).
\label{fig:deck2}}
\end{figure}

We begin with a synthetic example of a parabolic signal with random noise, see \cite{2021ApJS..252...11A} for a similar example with a linear synthetic signal with noise. In IDL, it can be created using the following commands,
\vspace{1mm} 
\begin{verbatim}
x = 0.1d*findgen(100)
noise = randomn(1,100)
noise /= stdev(noise)
y = 1d + 5d*x - 0.5d*x^2 + 2d*noise
\end{verbatim}
\vspace{1mm} 
The visualisation of such a synthetic signal is shown in Fig.~\ref{fig:mcmc_synth}. To fit this signal with SoBAT, we consider three guessed functions with linear, quadratic, and cubic dependence upon $x$, with two, three, and four free parameters, respectively. For example, one can use
\vspace{1mm} 
\begin{verbatim}
function cubic_model, x, params, _extra = _extra
  b = params[0]
  k = params[1]
  c = params[2]
  d = params[3]
  return,  b + k*x + c*x^2 + d*x^3
end
\end{verbatim}
\vspace{1mm} 
for creating the cubic model in IDL. To use those models for fitting our synthetic signal with SoBAT, one needs to provide a rough initial guess for the model parameters and specify the intervals of their expected values, based on prior expert knowledge (for example, we may demand the plasma parameter $\beta$ to be always positive and smaller than 1 in the corona). SoBAT allows for both the uniform and normal distributions of the model parameters within those intervals. For the above cubic model and uniform distribution of priors, this can be done in IDL as
\vspace{1mm} 
\begin{verbatim}
pars_cubic = [1d,1d,0.1d,0.1d]
priors_cubic = objarr(4)
priors_cubic[0] = prior_uniform(-5d,5d) ;prior_normal for normal
priors_cubic[1] = prior_uniform(-10d,10d)
priors_cubic[2] = prior_uniform(-1d,1d)
priors_cubic[3] = prior_uniform(-1d,1d)
\end{verbatim}
\vspace{1mm} 
Now we can use the above model functions and the prior guess for the values of their parameters to call the \verb|mcmc_fit| function within SoBAT software package for best-fitting our experimental signal $y(x)$. In IDL:
\vspace{1mm} 
\begin{verbatim}
fit_mcmc_cubic = mcmc_fit(x, y, pars_cubic, 'cubic_model',$
                 priors = priors_cubic, burn_in = 100000,$
                 n_samples = 1000000, samples = samples_cubic,$
                 credible_intervals = intervals_cubic,$
                 confidence_level = 0.95)
\end{verbatim}
\vspace{1mm} 
In this example, we request $10^5$ samples for preliminary sampling and $10^6$ samples for main sampling which is most often used in practice. Each of these samples represents a combination of free model parameters which provide an acceptable fit based on the Bayes probabilistic approach, see \cite{2021ApJS..252...11A} for more details. All $10^6$ successful samples (combinations of the model parameters) are stored in the \verb|samples_cubic| variable in the above example. After the main sampling has finished, the best-fitting combination of the model parameters is returned in the \verb|pars_cubic| variable, with the corresponding credible intervals (\verb|intervals_cubic|) estimated at the prescribed 95\% confidence level.

\begin{table}
	\centering
 \caption{The best-fit model parameters inferred with SoBAT for the synthetic signal $y(x)$ shown in Fig.~\ref{fig:mcmc_synth}, approximated by the linear ($b+kx$), quadratic ($b+kx+cx^2$), and cubic ($b+kx+cx^2+dx^3$) functions.}
 \label{tab:mcmc_synth}
 \renewcommand{\arraystretch}{1.5}
	\begin{tabularx}{0.79\textwidth}{XXXXX} 
        \hline
        \hline
    Model & $b$ & $k$ & $c$ & $d$  \\
        \hline
    linear& $8.8^{+1.7}_{-1.7}$ &  $0.1^{+0.3}_{-0.3}$     & -- & -- \\
    quadratic&  $0.7^{+1.1}_{-1.2}$   &  $5.1^{+0.5}_{-0.6}$ & $-0.5^{+0.05}_{-0.06}$ & -- \\
    cubic& $0.3^{+1.6}_{-1.4}$    &  $5.5^{+1.3}_{-1.4}$    & $-0.6^{+0.3}_{-0.3}$ & $0.007^{+0.02}_{-0.02}$ \\
    \hline
	\end{tabularx}%
\end{table}

Table~\ref{tab:mcmc_synth} shows the model parameters that we obtain for the synthetic signal shown in Fig.~\ref{fig:mcmc_synth} after best-fitting it with the linear, quadratic and cubic functions with SoBAT. Both quadratic and cubic models result in almost identical fits, well consistent with the input signal, and the estimated values of the parameters $b$, $k$, and $c$ for these models reproduce the input values within the obtained credible intervals. We note that if the obtained best-fit value of a model parameter appears to be close to the maximum or minimum value of the prior guess interval fixed above, the latter has to be broadened within reasonable limits to allow the \verb|mcmc_fit| function to scan through a broader range of this parameter.
Moreover, the credible intervals in Table~\ref{tab:mcmc_synth} are almost symmetric which is not always the case in practice, see, e.g., Figs. 5 and 6 in \cite{2022ApJ...929..101P}. In other words, SoBAT can readily handle cases with non-Gaussian distributions of the model parameters, which is another advantage of SoBAT in comparison with standard least-square methods.

The parameter $d$ in the cubic model is found to be practically zero and, hence, redundant. The linear model, in turn, is only able to capture the mean value of the input signal and thus is clearly insufficient. SoBAT allows us to perform a rigorous comparison between several competing models with the use of the \verb|mcmc_fit_evidence| function and calculation of the Bayes factor $K_{i,j}$ through the ratio of the Bayes evidence for models \lq\lq i\rq\rq\ and \lq\lq j\rq\rq. For the cubic model, for example, the Bayes evidence can be calculated in IDL as
\vspace{1mm} 
\begin{verbatim}
e_cubic = mcmc_fit_evidence(samples_cubic, x, y,$
          priors_cubic, 'cubic_model')
\end{verbatim}
\vspace{1mm} 
using the \verb|samples_cubic| variable obtained above with the \verb|mcmc_fit| function. Thus, having the Bayes evidence for the linear $B_\mathrm{lin}$, quadratic $B_\mathrm{quad}$, and cubic $B_\mathrm{cubic}$ models estimated, we can obtain their ratios as $B_\mathrm{quad}/B_\mathrm{lin} \approx 10^{31}$ and $B_\mathrm{quad}/B_\mathrm{cubic} \approx 61$. This means that the quadratic model is substantially better for our input signal than the linear model as the order of the latter is insufficient to describe the whole picture. Likewise, the quadratic model is more suitable than the cubic model as it has more free parameters one of which is found to be redundant and not playing a role in the physical process under analysis.

We now demonstrate the application of the pipeline described above to best-fitting the kink oscillation of a coronal loop shown in Fig.~\ref{fig:deck2} with several competing models. More specifically, an exponential damping model $M_\mathrm{e}(t)$, e.g., \cite{2006RSPTA.364..433G} and a generalised model (consisting of Gaussian and exponential damping patterns) $M_\mathrm{g}(t)$, e.g., \cite{2016A&A...585L...6P} are employed to fit the displacement signal (presented as red dots in Fig.~\ref{fig:deck2}) of a kink-oscillating loop.
Thus, the model function can be characterised as a decaying harmonic function superimposed on a background trend,
\begin{align}
    &\xi(t) = A M(t) \sin \left(\frac{2 \pi}{P} t+\varphi\right)  + T(t), \label{eq:signal}\\
    &M_\mathrm{e}(t)= \exp \left(-\frac{t}{\tau_\mathrm{e}}\right),\nonumber \\
    &M_\mathrm{g}(t)=\begin{cases}
    \displaystyle \exp \left(-\frac{t^2}{2 \tau_\mathrm{g}^2}\right),  & t \leq t_\mathrm{s}, \\ 
    \displaystyle A_\mathrm{s} \exp \left(-\frac{t-t_\mathrm{s}}{\tau_\mathrm{ge}}\right), & t>t_\mathrm{s}.\nonumber
    \end{cases}
\end{align}
where $A$ is the oscillation amplitude, $M(t)$ stands for different damping models, $P$ is the oscillation period, $\varphi$ is the initial phase, and $T(t)$ is a low-order polynomial trend.
In this example, we consider quadratic and cubic polynomial functions for the background trend $T(t)$ to study if the choice of a model function for $T(t)$ affects the kink damping model comparison.
In $M_\mathrm{e}(t)$ and $M_\mathrm{g}(t)$, $\tau_\mathrm{e}$ represents the exponential damping time, $\tau_\mathrm{g}$ and $\tau_\mathrm{ge}$ are the characteristic damping times of the Gaussian and exponential phases in the Gaussian--exponential model, respectively, and $t_\mathrm{s}$ is the switch time between them.

We use SoBAT to best-fit the observed kink oscillation for four possible scenarios: the exponential or Gaussian-exponential damping with the quadratic or cubic background trend (see panels (b) and (c) in Fig.~\ref{fig:deck2}). For this, we perform the steps described above, namely: function (\ref{eq:signal}) initialisation in IDL; providing reasonable initial guess and ranges for the model parameters; calling the \verb|mcmc_fit| and \verb|mcmc_fit_evidence| functions for best-fitting and model comparison.
The resulting best-fitting key model parameters and the corresponding Bayes factors are summarised in Table~\ref{tab:mcmc_kink_exa}.

For the quadratic and cubic polynomial background trends, the Bayes factor $K_{g,e}$ are found to be about 100 and 120, respectively. This result indicates that the input displacement signal strongly favours the interpretation by Model $M_\mathrm{g}(t)$ than by Model $M_\mathrm{e}(t)$, and the choice of the functional form of the background trend does not alter it in this particular case study. Indeed, the Gaussian-exponential model $M_\mathrm{g}(t)$ visually displays a better fit than the exponential model $M_\mathrm{e}(t)$ in both panels (b) and (c) of Fig.~\ref{fig:deck2}.
On the other hand, the Bayes factor $K_{cubic,quad}$ (about 68 for $M_\mathrm{e}(t)$ and 87 for $M_\mathrm{g}(t)$) suggests that the cubic function could be a preferred background trend in this case.
Thus, the blue best-fitting curve in Fig.~\ref{fig:deck2}(c) appears to provide the best match with the original signal among all four combinations considered.
We also note that, in general, models with more free parameters are penalised in Bayesian comparison due to the increased parameter space.

\begin{table}[h]
	\centering
	\caption{The best-fit model parameters and Bayes factors inferred with SoBAT for the kink oscillation event shown in Fig.~\ref{fig:deck2}, approximated by the exponential $M_\mathrm{e}(t)$ or Gaussian-exponential $M_\mathrm{g}(t)$ damping model with the quadratic $T_\mathrm{quad}(t)$ or cubic $T_\mathrm{cubic}(t)$ background trend (see Eq.~\ref{eq:signal}). Four possible scenarios are considered which are M1: $M_\mathrm{e}$ + $T_\mathrm{quad}$, M2: $M_\mathrm{e}$ + $T_\mathrm{cubic}$, M3: $M_\mathrm{g}$ + $T_\mathrm{quad}$, M4: $M_\mathrm{g}$ + $T_\mathrm{cubic}$.}
	\label{tab:mcmc_kink_exa}
 \renewcommand{\arraystretch}{1.5}

	\begin{tabular}{l|cccc|cccccc} 
        \hline
        \hline
     & \multicolumn{4}{c|}{Bayes factor} & Osc. Amp. & Period &      Exp.        & Gaussian-exp.              & \# of   \\
     & \multicolumn{4}{c|}{$K_{i,j}$ (M$_i$ vs. M$_j$)}  &    $A$, Mm      &   $P$, min    &     $\tau_e$, min     &  $\tau_g$ / $\tau_{ge}$, min    &  params     \\
        \hline
    M1  & -- & -$68$ & -$100$ & -- & $11.2^{+0.9}_{-1.0}$   &  $8.3^{+0.1}_{-0.1}$ & $18.0^{+2.7}_{-1.8}$  & --   & 7   \\
    M2  & $68$ & -- & -- & -$120$  & $12.6^{+0.9}_{-0.9}$   &  $8.1^{+0.1}_{-0.1}$ & $16.1^{+1.6}_{-1.5}$  & --   & 8   \\
    M3  & $100$ & -- & -- & -$87$ & $9.6^{+0.5}_{-0.6}$    &  $8.2^{+0.1}_{-0.1}$ & --  & $39.7^{+23.9}_{-9.6}$ / $5.5^{+3.1}_{-3.4}$  &  9 \\
    M4  & -- & $120$ & $87$ & -- & $10.1^{+0.5}_{-0.4}$   &  $8.2^{+0.1}_{-0.1}$ & --  & $37.5^{+26.4}_{-7.2}$ / $3.7^{+4.0}_{-1.6}$  &  10 \\
    \hline
    &  M1 &  M2 &  M3 &  M4  &&&&& \\
	\end{tabular}
\end{table}

\subsection{Differential emission measure analysis}
\label{sec:dem}
\begin{figure}
    \centering
    \includegraphics[width=0.95\columnwidth]{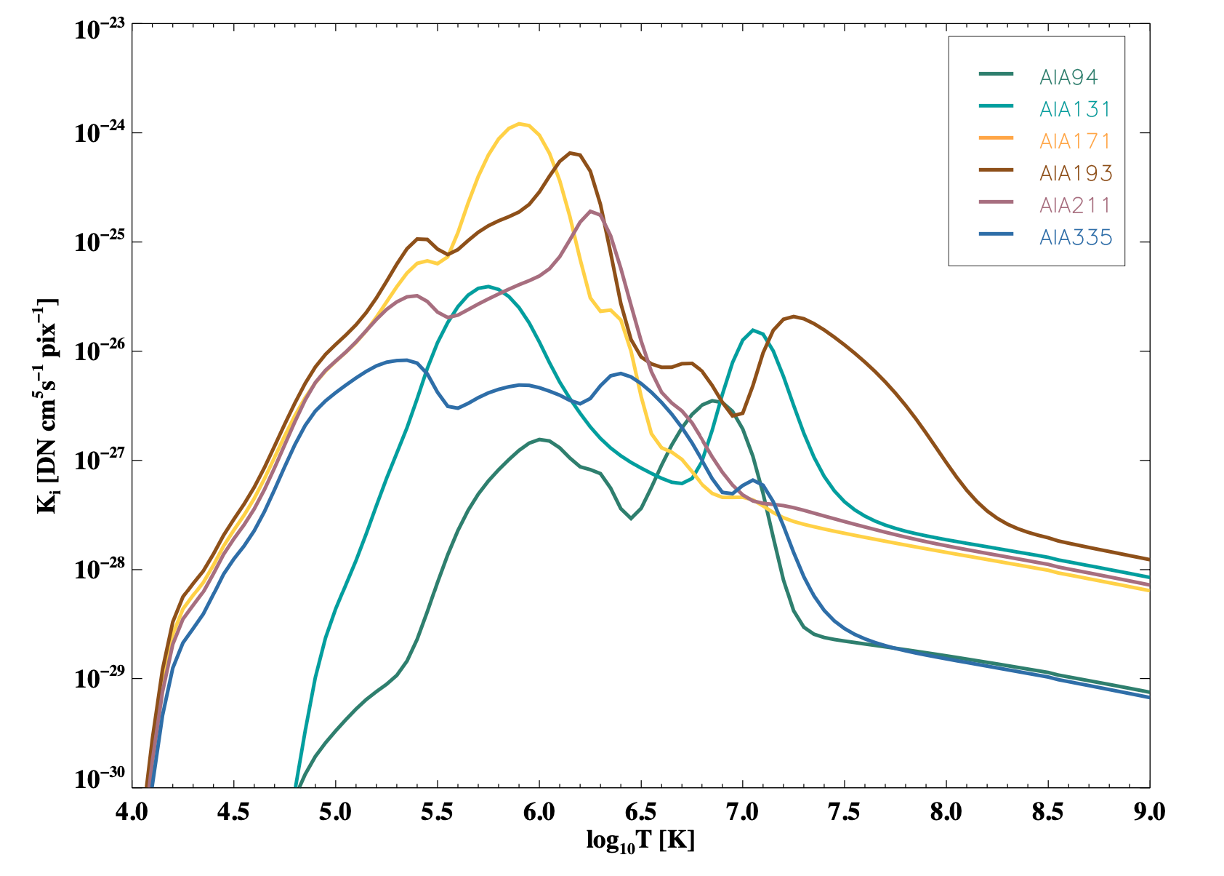}
    \caption{Temperature response function of six EUV channels of SDO/AIA. The corresponding EUV wavelengths are indicated in the legend. 
    \label{fig:tresp}}
\end{figure}

\begin{figure}
    \centering
    \includegraphics[angle=270,width=\columnwidth]{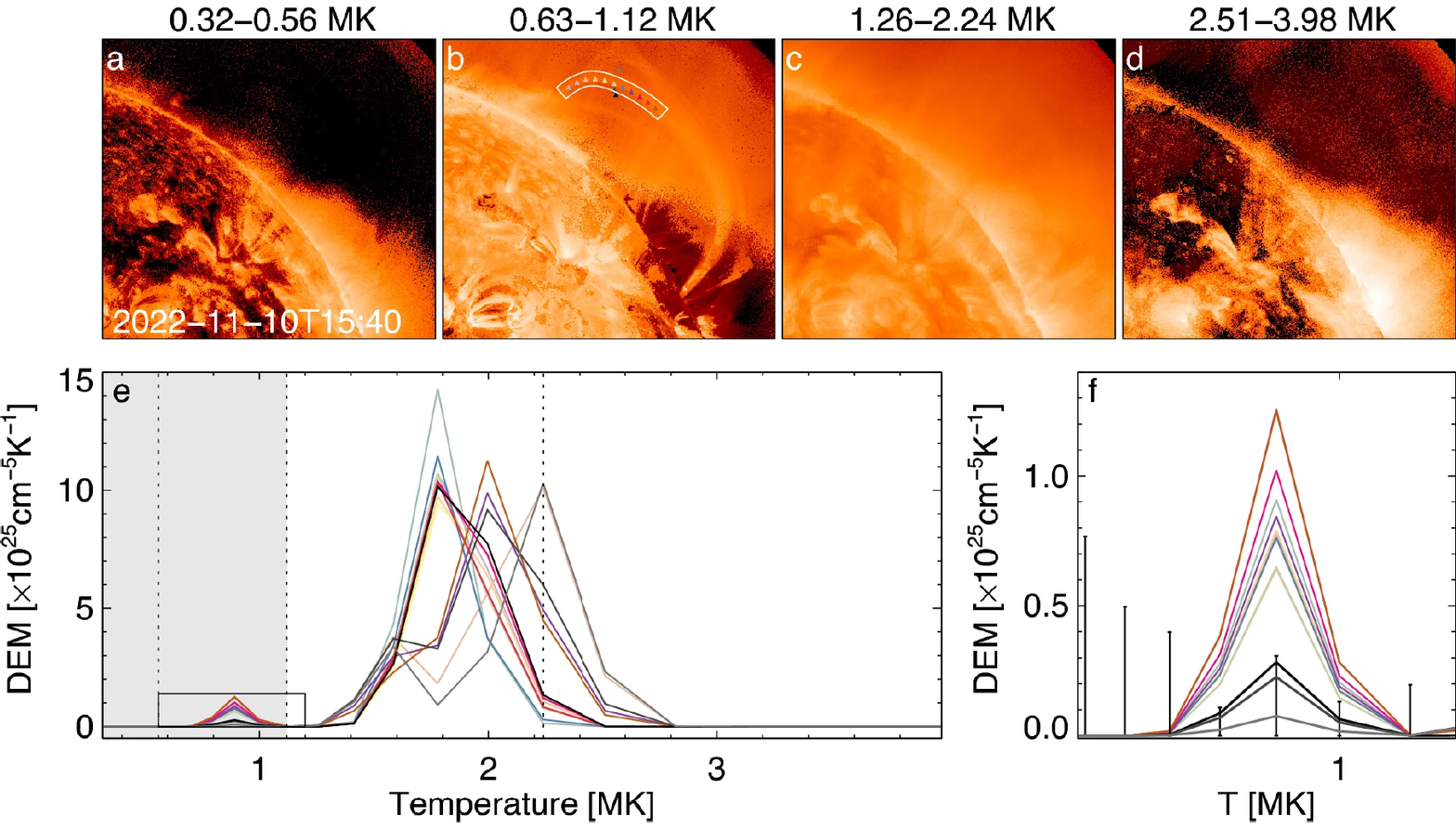}
    \caption{An example of DEM maps for various temperature ranges and DEM profiles {(panel e)} for selected pixels {(marked by the coloured triangles) inside and outside the analysed loop. The white slice in panel b highlights the loop segment which displays decayless kink oscillations, and this segment is used for plasma diagnostics. Panel f is a zoom-in version of the small window in panel e. The vertical dotted lines divide the temperature range into 4 subsets as shown in panels a–d.} Used with permission from \cite{2023NatSR..1312963Z}.
    \label{fig:DEM}
    }
\end{figure}

The temperature and density of the solar plasma are the vital parameters for performing coronal MHD seismology.
In the corona, the 
observed 
EUV flux could contain the emission from multi-thermal structures situated along the LoS. The EUV radiation flux is directly proportional to the emission measure (EM), which is defined as the quantity of the plasma integrated along the LoS per unit area, i.e., the pixel, under the optically thin assumption, by
\begin{equation}
   \mathrm{EM}=\int_z n_e^2(z) dz = \int_T \xi(T) dT, \label{eq:em} 
\end{equation}
where $n_e$ is the number density of electrons and $z$ is the coordinate along LOS. The quantity EM is measured in units of cm$^{-5}$. As EM is the integrated emission flux along LoS and does not reflect the thermal structure of the coronal plasma, one would be more interested in getting the differential emission measure (DEM, $\xi$, measured in cm$^{-5}$K$^{-1}$), which is defined by 
\begin{equation}
   \xi  =n_e^2(T) \frac{dz}{dT}. \label{eq:dem} 
\end{equation}
Here we should note the mapping between $z$ and $T$, $n_e(T)=n_e(T(z))=n_e(z)$, the factor ${dz}/{dT}$  arises from the change of variable from $z$ to $T$ in integration in Equation~\ref{eq:em}. 

In EUV observations, such as SDO/AIA, the measured value at each pixel $y_i \,\mathrm{[DN\, s^{-1}]}$ in the $i^{th}$ AIA channel is related to the DEM by a temperature response function $K_i(T) \mathrm{[DN cm^5 s^-1]}$ and is defined as
\begin{equation}
    y_i=\int_T K_i(T) \xi(T) dT.\label{eq:aia}
\end{equation}
This is the theory for the inversion of DEM by narrow-band EUV imaging observations. The temperature response functions for the six AIA channels are presented in Figure~\ref{fig:tresp}. 

Equation~\ref{eq:aia} is assessed in discrete form. As AIA has six EUV channels, so we set the number of channels to $M=6$. We could opt to assess DEM with $N=20$ temperature bins, say from $\log T = 5.5$ to $\log T =7.5$ with a step of $\Delta \log T=0.1$, such that the temperature grid is uniform in logarithmic scale. In practical use, the number of temperature grid $N$ is greater than the number of channels $M$, Equation~\ref{eq:aia} is under-determined, so we have to resort to a mathematical model for $\xi(T)$ with a few free parameters. We then evaluate DEM at temperature $T_j$, $\xi(T_j)=\xi_j$, and minimize the following term with respect to the free parameters in the mathematical model of $\xi(T)$, 
\begin{equation}
    \text{arg min} \left|\frac{\mathbf{K}\mathbf{E}-\mathbf{Y}}{\delta\mathbf{Y}}\right|^2 \label{eq:argmin}
\end{equation}
where $\mathbf{Y} \in \mathcal{R}^{M}$ is a column vector formed by EUV emission intensity measured with $M$ AIA channels, $\delta\mathbf{Y}$ is the associated uncertainty; $\mathbf{K}\in \mathcal{R}^{M\times N}$ with $K_{ij}=K_i(T_j)$; $E_j = \xi_j\Delta T_j$ is product of the DEM at $T_j$ and the temperature interval between two adjacent temperature grid points. Equation~\ref{eq:argmin} is normally solved with forward modelling \cite{2013SoPh..283....5A} or regularization \cite{2012A&A...539A.146H,2015ApJ...807..143C,2018ApJ...856L..17S}. 

To determine DEM, we need to solve Equation~\ref{eq:aia} inversely, based on the observed intensity in each channel. With the inverted DEM model, we could readily calculate the plasma temperatures, by  
\begin{equation}
T_{\mathrm{DEM}}=\frac{\sum\xi_j \Delta T_j T_j}{\sum \xi_j \Delta T_j}. \label{eq:T_dem}
\end{equation}
We shall note that this temperature is the DEM-weighted temperature and acts as an estimate of the temperature of a {coronal structure. However, this approach should be used with a caution, as coronal plasma structures could consists of layers with rather different temperatures.} 

Figure~\ref{fig:DEM}{a-d} shows a typical example of DEM inversion and calculation of plasma temperature and density. A giant coronal loop was detected above the solar limb (Figure~\ref{fig:DEM}{b}), which is also filled with stratified coronal plasma (Figure~\ref{fig:DEM}{c}). We performed DEM inversion and extracted the DEM profiles of several sample pixels located on the coronal loop. The DEM profiles ( Figure~\ref{fig:DEM}{e}) have two peaks, a major one at about 2 MK and a minor one at about 0.9 MK. So we infer that the two peaks record the plasma emission from the coronal loop (Figure~\ref{fig:DEM}{b}) and the diffuse coronal background (Figure~\ref{fig:DEM}{c}). {Here, it is assumed that the background density is constant along the LoS and the column depth is calculated according to the geometrical model of a stellar atmosphere \citep{1936HarCi.417....1M,2014A&A...564A..47Z}.} So we took a comparative DEM profile (black line) from a pixel at the coronal background without the coronal loop. This DEM profile exhibited a major emission peak at about 2 MK, however has negligible emission at 0.9 MK. With this comparison, we conclude that the coronal loop is a uni-thermal coronal structure with plasma temperature of about 0.9 MK. We then calculated the DEM-weighted temperature (Equation~\ref{eq:T_dem}) within the range of 0.32 MK to 1.12 MK and obtained a temperature of 0.88 MK for the coronal loop. We subsequently calculated the EM of the coronal loop by integrating the DEM of the minor peak (0.32 MK to 1.12 MK) and subtracting the background emission and calculated the density of the coronal loop by $n_e=\sqrt{EM/w}$, where $w$ is the geometric width of the coronal loop. The number density of electrons of the coronal loop measured at multiple locations is $0.9^{+0.4}_{-0.2}\times10^8\mathrm{cm}^{-3}$. These parameters obtained with DEM inversion are then used in coronal MHD seismology to calculate magnetic field strength. 

The code used to calculate the DEMs in the above is the Sparse inversion code (written in IDL, \citep{2015ApJ...807..143C}) with improved settings \citep{2018ApJ...856L..17S}. The latest version is available at {\url{http://paperdata.china-vo.org/yang.su/DEM/sparse_em_v1.001_ys.zip}} \citep{2022ApJ...930..147L}, which includes an example script demonstrating how to obtain EM(T) images from AIA data. Here's a quick overview of the process: \\
\begin{enumerate}
\item Start by downloading images in six EUV bands at the target time instance and apply exposure normalisation.

\item Save the six normalized images as a 3D datacube: \verb|map_in = [nx, ny, wavelength]|. This datacube serves as the input AIA data for the main procedure, \verb|get_em_from_aia.pro|. If your target field of view is large, it is recommended to reduce the data size by setting the parameter n\_pix to rebin n\_pix $\times$ n\_pix pixels as one.

\item Determine all the necessary parameters for the main procedure. To obtain uncertainty estimates for the EM results, set n\_mc, the number of Monte-Carlo (MC) simulation runs.

\item Run \verb|get_em_from_aia.pro|, and it will return a 4D array, \verb|[nx, ny, nT, n_mc+1]|. In this array, each pixel contains DEM in unit of $\mathrm{cm^{-5}K^{-1}}$ in predefined temperature bins and for each MC run.
\end{enumerate}

\section{Example recipes of seismological inversions}
\label{sec:example}

In this section we give several practical recipes for seismological diagnostics of coronal plasma structures. 

\subsection{Estimating the magnetic field in coronal loops by kink oscillations}
\label{sec:magseis}

The dependence of kink oscillations of coronal loops on the kink speed allows us to estimate the absolute value of the magnetic field in those plasma structures. In the standard form, this seismological technique is based upon the assumption that the magnetic field along the loop is constant, which is consistent with observations which show that a loop's width does not change with height. In addition, it is assumed that the plasma density is constant along the loop and in the external plasma, i.e., that the effect of stratification can be neglected. Thus, the kink speed along the loop remains constant. {Additionally, there are other simplifying assumptions, including the linear wave regime, uniform temperatures inside and outside the waveguide, and no flows in the waveguide.}

\begin{itemize}
\item The first step is to establish that the repetitive kink displacement of the loop of interest is a standing wave. It is demonstrated by the in-phase or anti-phase nature of the displacements of different segments of the loop. If the phase behaves differently, the wave is propagating, and its period is prescribed by the driver. 
\item The oscillation period and the wavelength of the standing oscillation should be determined independently.
The oscillation period is determined with the use of a time--distance map of a kink oscillation either by a visual inspection, or fitting the oscillatory pattern by a guessed function, for example, using the MCMC technique described in Section~\ref{sec:mcmc}. The latter approach is recommended in the cases when the oscillatory pattern is suspected to be a superposition of several oscillatory harmonics.  
\item Assuming that the loop's shape is approximately half-circular, the loop length could be estimated as the distance between the footpoints, $d_\mathrm{fp}$, as $L = \pi d_\mathrm{fp}$. The distance $d_\mathrm{fp}$ is measured as the chord length between the Stonyhurst-Heliographic (HG) coordinates of the footpoints.
One can convert the native solar coordinates of a FITS file using the Word Coordinate System into the HG system, by the \verb|wcs_convert_from_coord.pro| {IDL function from SSW library.}
With the HG coordinates of two foopoints: $[\lambda_1, \phi_1]$ for footpoint1, $[\lambda_2, \phi_2]$ for footpoint2, where $\lambda$ is longitude and $\phi$ is latitude, one can calculate the angle between the two footpoints on the sphere by 
\begin{equation}
    \theta = \arccos(\cos \phi_1 \cos \phi_2 \cos\left( |\lambda_1 - \lambda_2|\right) + \sin \phi_1 \sin \phi_2 )
\end{equation}

Then the arc length $c$ connected the two footpoints on the sphere (also called great-circle distance) is obtained $c = \theta R_\odot $, and the chord length $d_\mathrm{fp}$ is calculated as $ 2 R_\odot \sin(\frac{\theta}{2})$.
Alternatively, the chord length could be attempted by converting the point from sphere $[ R_\odot,\lambda, \phi] $ to Cartesian $[ R_\odot\sin\phi \cos\lambda, R_\odot\sin\phi \sin\lambda, R_\odot\cos\phi ]$ and then using Pythagoras. 

\item The determination of the wavelength requires also the identification of the parallel harmonic number of the kink mode of interest, which is calculated at the number of oscillation nodes, $N_\mathrm{nd}$, along the loop, excluding the naturally existing nodes at the footpoints. Then, the wavelength is $\lambda = 2L/(N_\mathrm{nd}+1)$.

\item Applying the DEM inversion technique (Section~\ref{sec:dem}), we estimate the plasma densities inside and outside the loop, $\rho_\mathrm{in}$ and $\rho_\mathrm{ex}$, respectively. In this estimation, we assume that the loop width along LoS is about its width in the PoS. 

\item 
The strength of the magnetic field in the loop is thus 
\begin{equation}
\label{eq:b0}
B \approx \frac{L}{P_\mathrm{kink} (N_\mathrm{nd}+1)} \Big[2\mu_0 \rho_\mathrm{in}\Big(1+\frac{\rho_\mathrm{ex}}{\rho_\mathrm{in}}\Big)\Big]^{1/2}.
\end{equation}
\end{itemize}

If several parallel harmonics are confidently detected, the field estimation could be improved by averaging the values obtained separately by different harmonics. However, one needs to be cautious with this step, as it requires the kink speed to be constant along the loop. If this condition is not fulfilled, for example, in larger loops with heights comparable to the stratification scale height, the oscillation periods of different parallel harmonics are determined by the values of the kink speed at different segments of the loop (e.g., \cite{2009SSRv..149....3A}). This effect manifests as the violation of the expected relationship between the periods of different harmonics, i.e., $P_\mathrm{kink}^{(1)}/( N_\mathrm{nd} P_\mathrm{kink}^{(N_\mathrm{nd}+1)}) \neq 1$, where the superscripts indicate the harmonic number. Moreover, the ratio $P_\mathrm{kink}^{(1)}/( N_\mathrm{nd} P_\mathrm{kink}^{(N_\mathrm{nd})})$ provides important seismological information about the density scale height, and can also be used for probing the variation of the magnetic field along a loop with a  non-constant minor radius \cite{2008A&A...486.1015V}. 

In the rapidly decaying regime, kink oscillations could also be used for probing the fine, sub-resolution structuring of the coronal plasma across the field. The specific parameter estimated seismologically is the steepness of the perpendicular profile of the Alfv\'en speed by the exponential damping time, see Eq.~\ref{eq:td} \cite{2003ApJ...598.1375A}. A more sophisticated approach is based on the theoretically predicted departure of the damping pattern from the exponential, see \cite{2018ApJ...860...31P, 2019FrASS...6...22P} for detailed discussion, and also the example in Section~\ref{sec:mcmc}.

\subsection{Probing the coronal heating function}
\label{sec:heatseis}

Revealing the coronal heating mechanism remains one of the major outstanding problems in modern solar physics. Traditionally, MHD waves have been considered as potentially responsible for heating the corona via the transfer of energy from the lower atmosphere upwards. However, the most recent results converge to the conclusion that the observed wave amplitudes and propagation speeds are not sufficient to compensate for colossal energy losses from the {coronal active regions} via optically thin radiation and thermal conduction towards the cooler chromosphere \citep{2020SSRv..216..140V}. In more recent years, a new approach to the problem of MHD waves and coronal heating has been proposed, in which the waves are considered as not the cause of the coronal plasma heating but natural probes of the heating process via the phenomenon of wave-induced thermal misbalance \citep{2021PPCF...63l4008K}. Indeed, due to the continuous interplay between plasma cooling and heating processes in the corona, essentially compressive MHD waves perturb not only the mechanical equilibrium (i.e. the local force balance) but also the thermal equilibrium (i.e. the local thermal balance). In this section, we describe how the coronal heating function can be probed with the theory of thermal misbalance and observations of decaying slow magnetoacoustic waves in long-lived thermodynamically stable coronal plasma structures (mainly following \citep{2020A&A...644A..33K} and \citep{2022MNRAS.514L..51K}).

\begin{itemize}
\item As the observed lifetime of coronal loops is typically longer than the characteristic slow-wave oscillation period or damping time (from a few to a few tens of minutes, see Secs.~\ref{sec:slow} and \ref{sec:sumer}), we begin with the assumption that the wave-hosting loop remains stable to perturbations of the local thermal equilibrium, i.e. no rapid plasma condensations such as coronal rain is developed (cf., \citep{2022FrASS...920116A}).

\item Another assumption is to parametrise the unknown coronal heating function in a generic power-law form via local values of the macroscopic coronal plasma parameters, i.e. density, temperature, and magnetic field, $H(\rho, T, B) \propto \rho^aT^bB^c$. In the limit of low plasma-$\beta$, slow waves cause very small perturbation of the loop's magnetic field (e.g., \citep{2022ApJ...926...64O}), so that the dynamics of slow waves becomes practically insensitive to the power-law index $c$ (e.g., \citep{2021A&A...646A.155D}). The latter allows us to reduce the number of unknown parameters from three, $(a,b,c)$ to two, $(a,b)$.
{For finite-$\beta$ regimes, such as in hot and dense flaring loops (see e.g. \citep{2013ApJ...779L...7K}), one should use the generalised approach described in \citep{2023Physi...5..193K}.}

\item The thermodynamic stability of the coronal loop to acoustic and entropy modes (also known as isentropic and isobaric or thermal, respectively) is wavelength-dependent. In this respect, it is useful to consider the so-called acoustic and entropy Field's lengths,
\begin{align}
    &\lambda_\mathrm{F}^\mathrm{acoustic} = 2\pi \sqrt{\frac{\kappa_\parallel T_0}{\rho_0 L_0 \left[ \frac{a-1}{\gamma - 1} + b - \frac{T_0}{L_0}\frac{\partial L_0}{\partial T}\right]}}, \label{eq:thermal_instab_ac}\\
    &\lambda_\mathrm{F}^\mathrm{entropy} = 2\pi \sqrt{\frac{\kappa_\parallel T_0}{\rho_0 L_0 \left[ 1-a+b-\frac{T_0}{L_0}\frac{\partial L_0}{\partial T}\right]}}, \label{eq:thermal_instab_ent}
\end{align}
which represent the characteristic wavelengths above which the parallel thermal conduction becomes insufficient to counteract the instability of acoustic and entropy perturbations.
Evaluating the instability conditions (\ref{eq:thermal_instab_ac}) and (\ref{eq:thermal_instab_ent}) for the equilibrium coronal plasma density $\rho_0$, temperature $T_0$, optically thin radiation function $L_0$ taken, for example, in the CHIANTI form \citep{2021ApJ...909...38D}, field-aligned thermal conduction coefficient $\kappa_\parallel$ taken, for example, in the Spitzer form ($\propto T_0^{5/2}$), standard adiabatic index $\gamma=5/3$, and the perturbation's wavelength $\lambda$, one can derive the parametric region in the $(a,b)$-plane outlining the heating models which allow for a long-lived thermodynamically stable coronal loop. 
We also note that the local Spitzer approximation for $\kappa_\parallel$ was shown to break down for the perturbation's wavelength less than 5\,Mm and 500\,Mm for a 1-MK and 10-MK coronal plasma, respectively, and the use of non-local transport models or thermal flux limiters is required \citep{2023FrASS..1055124A}.

\item One can further refine the region of $(a,b)$ parameters, derived at the previous step, by accounting for the observed frequency-dependent damping of slow waves. Indeed, the relationship between the observed oscillation period and damping time of slow waves in coronal loops is strongly affected by non-adiabatic processes such as thermal conduction and thermal misbalance. By solving the general dispersion relation $D(\omega^3,\omega^2,\omega,k^4,k^2,\kappa_\parallel,a,b)=0$ (e.g., \citep{2021SoPh..296...20P}) either numerically or analytically, one can obtain combinations of the heating parameters $(a,b)$ within the above-derived stability region, which are consistent with both the observed oscillation period and damping time (or damping length for propagating waves). In a weakly non-adiabatic and low-$\beta$ regime, the approximate relationship between the standing slow wave damping time $\tau_\mathrm{D}$ and oscillation period $P$ takes the form
\begin{equation}\label{eq:freq-per_slow}
    \tau_\mathrm{D} = \frac{2\tau_\mathrm{M}P^2}{d\tau_\mathrm{M} + P^2},
\end{equation}    
\begin{equation}
\mathrm{where}~~
d=\frac{4\pi^2(\gamma-1)\kappa_\parallel}{\gamma\rho_0 C_\mathrm{V}c_\mathrm{s}^2},~~
\tau_\mathrm{M}=\frac{\gamma}{\gamma-1}\frac{C_V}{\dfrac{\partial \mathcal{L}_0}{\partial T}-\dfrac{\mathcal{L}_0}{T_0}\left(\dfrac{a-1}{\gamma-1}+b\right)},\nonumber
\end{equation}
which accounts for a combined effect of parallel thermal conduction and thermal misbalance as two major slow-wave damping mechanisms in the corona; $C_V = {(\gamma - 1)^{-1}k_\mathrm{B}}/{m}$ and $c_\mathrm{s}=\sqrt{{\gamma k_\mathrm{B}T_0}/{m}}$ are specific heat capacity and standard sound speed with Boltzmann constant $k_\mathrm{B}$ and the mean particle mass $m=0.6m_\mathrm{p}$, respectively. It is also instructive to use the Bayes analysis with MCMC approach described in Sec.~\ref{sec:mcmc} to perform a rigorous comparison of different damping scenarios of slow waves in the corona and assessment of a functional form of the heating process \citep{2023A&A...677A..23A}.
\end{itemize}

The above steps allow for estimating the steady-state uniform coronal heating function \citep{2023ApJ...957...25J} or if its intermittent nature is effectively averaged over the slow wave oscillation period. For probing the duration and spatial location of impulsive heating events, one can use observations of slow waves in the form of quasi-periodic pulsations in flares \citep{2019ApJ...884..131R}. 

\subsection{Determining the effective adiabatic index and transport coefficients by slow waves}
\label{sec:gamma}

The weak non-adiabatic limit is often assumed in MHD seismology studies as it simplifies the analysis. In this limit, energy exchange and transfer processes are weak and slow compared to typical time scales, such as the wave period. However, in the corona of the Sun, the weak non-adiabatic assumption does not necessarily hold as the observed wave damping times are often comparable to oscillation periods. In particular, non-ideal effects modify the slow wave phase speed ($V_\mathrm{ph}$) so that it can be expressed in a generic form as:
\begin{equation}
    V_\mathrm{ph}^2 = \left(\frac{\omega}{k}\right)^2 = \gamma_\mathrm{eff}(\kappa_\parallel, \eta, Q_{\rho,T,B}) \frac{c_\mathrm{s}^2}{\gamma},
\end{equation}
where $c_\mathrm{s}$ is the sound speed, $\gamma=5/3$ is the standard adiabatic index, and $\gamma_\mathrm{eff}(\kappa_\parallel, \eta, Q_{\rho,T,B})$ is the effective adiabatic index, which is dependent on non-adiabatic effects such as field-aligned conductivity ($\kappa_\parallel$), viscosity ($\eta$), and wave-induced perturbations of the coronal heating/cooling function and the effect of thermal misbalance ($Q_{\rho,T,B}$). Rearranging this equation gives us the effective adiabatic index:
\begin{equation}\label{eq:gamma_eff}
    \gamma_\mathrm{eff}= \gamma\left(\frac{V_\mathrm{ph}}{c_\mathrm{s}}\right)^2.
\end{equation}
That is, $\gamma_\mathrm{eff}$ characterises the deviation of the observed slow wave phase speed from the standard sound speed, caused by non-adiabatic effects in the corona\footnote{For standing waves, for which the wavelength is fixed by boundary conditions, $\gamma_\mathrm{eff}$ (\ref{eq:gamma_eff}) can be re-written through the observed oscillation period.}.  Hence, to calculate $\gamma_\mathrm{eff}$:
\begin{itemize}
    \item Determine the phase speed and predict the sound speed through other independent techniques. Then $\gamma_\mathrm{eff}$ can be calculated using Eq.~(\ref{eq:gamma_eff}). However, it is difficult to determine the absolute phase speed due to combined effects from projection and non-adiabatic processes that both modify the observed wave speed. 
    \item To cope with difficulties measuring the phase speed, additional observable parameters can be used, such as the ratio of relative temperature and density perturbations, $A_T/A_\rho$ (determined using DEM analysis, see Sec.~\ref{sec:dem}), and a phase shift between them, $\Delta \varphi$, in a slow wave. The polytropic assumption ($p\propto\rho^{\gamma_\mathrm{eff}}$) has regularly been used to determine $\gamma_\mathrm{eff}$ via $A_T/A_\rho$ in propagating, e.g., \cite{2018ApJ...868..149K} and standing, e.g., \cite{2015ApJ...811L..13W} slow waves, which assumes that:
\begin{equation}
    \gamma_\mathrm{eff}=\frac{A_T}{A_\rho}+1.
\end{equation}
 However, the polytropic assumption was shown only to be valid if non-adiabatic processes in the corona are significantly suppressed so that one can neglect the observed phase shift $\Delta \varphi$ and the expected deviation of $\gamma_\mathrm{eff}$ from $\gamma=5/3$ is small \citep{2022FrASS...973664K}. Thus, the polytropic assumption should be used with caution.
\item For more realistic coronal conditions, with full thermal conductivity as the major slow wave damping mechanism, reduced analytical solutions for $A_T/A_\rho$ and $\Delta \varphi$ take the form:
\begin{equation}\label{eq:AtAp}
    \frac{A_T}{A_\rho} \approx \frac{(\gamma -1)\cos\Delta\varphi}{1-2\pi\gamma d \chi(\gamma/\gamma_\mathrm{eff})},
\end{equation}
\begin{equation}\label{eq:tanphi}
    \tan\Delta\varphi \approx \frac{2\pi (\gamma -1)\kappa_\parallel m}{k_\mathrm{B}c_\mathrm{s}^2P_0\rho_0},
\end{equation}
where the thermal conduction parameter $d \propto \kappa_\parallel$ is given in Eq.~(\ref{eq:freq-per_slow}) and $\chi \equiv P/2\pi\tau_\mathrm{D}$ represents the effective oscillation quality factor, $\tau_D$ is the observed exponential damping time, $P$ is the observed wave period, and $P_0$ is the slow wave oscillation period in the ideal adiabatic case. Hence, we have two equations with two unknowns ($\kappa_\parallel$ and $\gamma_\mathrm{eff}$), and several other quantities that can be determined with independent observations. 
For example, in the regime of low thermal conductivity, $\gamma_\mathrm{eff}\to\gamma$, $A_T/A_\rho \to 2/3$, and $\Delta \varphi \to 0$. In an isothermal regime with very effective thermal conductivity,  $\gamma_\mathrm{eff}\to 1$, $A_T/A_\rho \to 0$, and $\Delta \varphi$ tend to some finite value below $\pi/2$. Interestingly, $\Delta \varphi$ caused by thermal conduction remains smaller than $\pi/2$ for both the fundamental and higher harmonics of a standing slow wave \citep{2023FrASS..1067781Z}, which provides another important observational constraint, see, e.g. \cite{2019MNRAS.483.5499K}.
\item Variations of $\gamma_\mathrm{eff}$ in a broader range are also possible if one takes other non-adiabatic effects into account, see, e.g., \cite{2019PhPl...26h2113Z} for the effect of thermal misbalance.

\item {The above approach can be generalised to probe the effect of compressive viscosity using observations and modelling of slow magnetoacoustic waves too (see a series of works \cite{2015ApJ...811L..13W, 2018ApJ...860..107W, 2019ApJ...886....2W, 2022ApJ...926...64O}). In particular, it was determined that the field-aligned heat conduction and viscosity coefficients in hot coronal loops need to be suppressed by a factor of 3 and enhanced by a factor of 10--15, respectively, to match observations.}

\end{itemize}



\subsection{Probing the direction of the coronal magnetic field}
\label{sec:bdir}

A key feature of slow waves is that they propagate almost parallel to the magnetic field in a low-$\beta$ plasma, such as the corona. Hence, if we can determine the wave vector of a propagating slow wave, we can infer the local direction of the magnetic field. This technique requires simultaneous observation from at least two instruments (quasi-stereoscopy) with non-parallel lines of sight.  The steps involved are as follows: 
\begin{enumerate}
    \item Produce time--distance maps as detailed in Sec.~\ref{sec:tdm} for each instrument. The slits should start where the slow waves first emerge and run parallel to the slow wave propagation direction. The maps should show diagonal ridges similar to those in Fig.~\ref{fig:Slow_Waves} (c).
    \item Determine the phase speeds, $V_\mathrm{ph}^\mathrm{(1,2)}$, for each instrument by determining the gradient of the ridges in the time distance map. These phase speeds are the speeds projected to each of the instrument PoS, hence both will be smaller than the actual phase speed, reduced by a factor of $\cos \theta$ where $\theta$ is the angular separation between the wave vector and the instrument PoS.
    \item By estimating the temperature of the plasma as the peak in the instrument response functions, we can approximate the actual phase speed, $V_\mathrm{ph}$, as the sound speed, given by
    \begin{equation}
        V_\mathrm{ph}(\mathrm{km\,s^{-1}}) \approx c_\mathrm{s}(\mathrm{km\,s^{-1}}) \approx 152 \sqrt{T\mathrm{(MK)}}.
    \end{equation}
    \item Calculate the angle between each instrument's wave vector and PoS using that $\theta_{1,2} = \cos ^{-1}{(V_\mathrm{ph}^\mathrm{1,2}/V_\mathrm{ph})}$. 
    \item $\theta_1$ and $\theta_2$ give four cones where the wave vector can lie, corresponding to angles $\pm \theta_\mathrm{1,2}$ from each PoS.  Where these cones intersect reduces the solution to four directions when using only two instruments. Some aspects of ``common sense" are then required to determine which direction is correct. For example, if the feature that hosts the propagating slow waves points eastwards in images from both instruments, westward-pointing solutions can be neglected. A third instrument would eliminate this need for manually disregarding solutions, and the point where all six cones intersect would give the wave vector.
\end{enumerate}

This technique would ideally use three high-resolution imaging instruments with different LoS. However, this configuration of instruments has not been available so far, except in the quasi-stereoscopic case when the three LoS were in the same plane (the ecliptic plane, e.g., \citep{2009ApJ...697.1674M}). A second downfall of this method is the determination of the temperature. Instrument response functions are broad; hence, the signal produced could be from plasma at any temperature within the response function. Thus, the temperature estimation is not exact. In addition, the polytopic index may differ from 5/3 due to non-adiabatic processes such as thermal conduction and thermal misbalance intrinsic for the corona (e.g., \citep{2019PhPl...26h2113Z}). 

Other approaches for probing the direction of the coronal magnetic field include: 
\begin{itemize}
    \item Dynamic stereoscopy (e.g., \citep{2015SoPh..290.2765A}) - this technique utilises the rotation of the Sun during long-duration observations to view a feature at many different angles. Dynamic stereoscopy requires that the feature does not develop throughout the observation, which can be as long as several days, and hence, it is difficult to use this method for more dynamic regions.
    \item Extrapolation (e.g., \citep{2012LRSP....9....5W}) - the magnetic field at the photosphere or chromosphere is extrapolated to produce a 3D magnetic geometry. This method is limited away from the centre of the solar disk, where observations of magnetic fields are impacted due to projection effects. 
    \item Propagating fast wave trains (see Sec. \ref{sec:qfp} and \ref{sec:qfpseis}) - the guided part of fast wave trains can be used analogously to propagating slow waves detailed above. 
\end{itemize}

\subsection{Probing fine cross-field structuring of the coronal plasma with quasi-periodic fast propagating wave trains}
\label{sec:qfpseis}
\begin{figure}
    \centering
    \includegraphics[width=0.95\columnwidth]{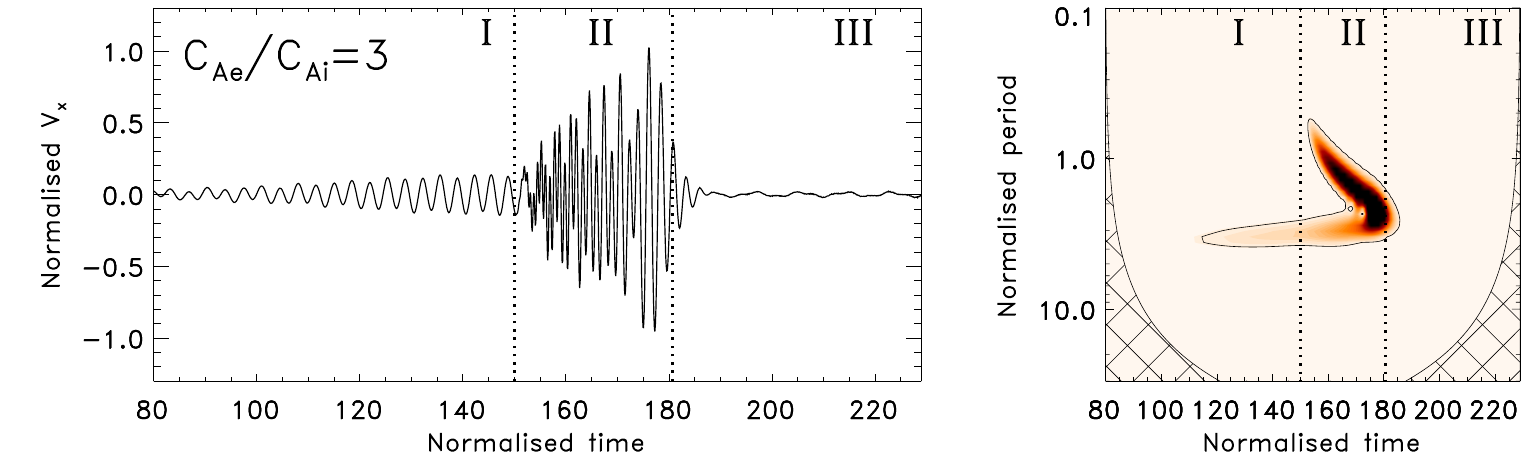}
    \includegraphics[width=0.95\columnwidth]{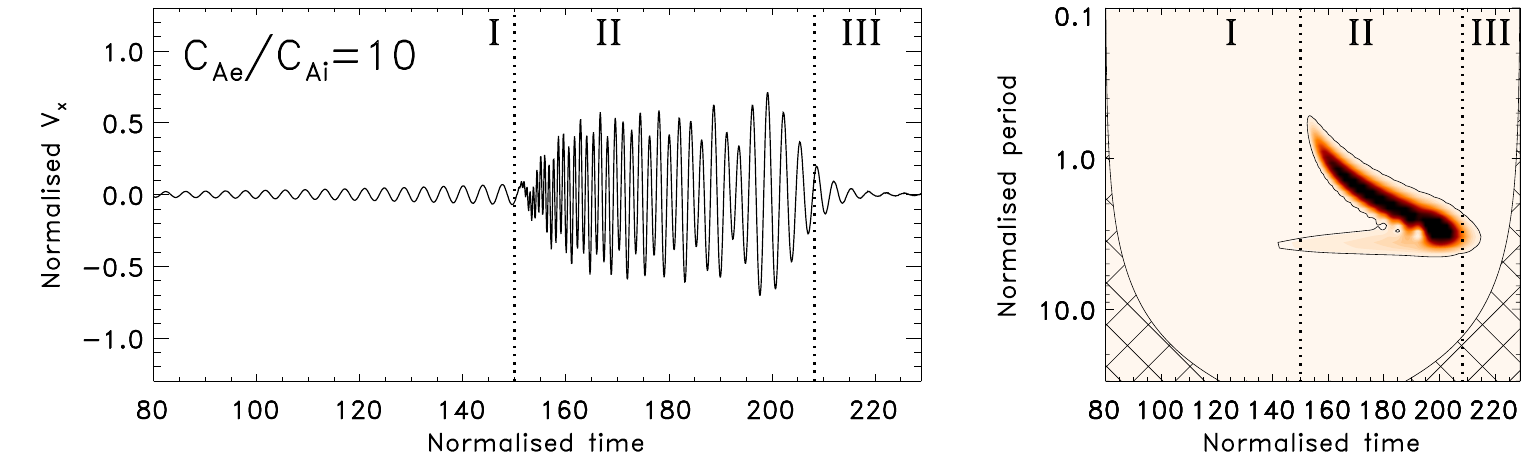}
    \caption{Fully developed QFP wave trains and the corresponding wavelet power spectra, modelled in the plasma slab with a step-function transverse density profile. For modelling technique and normalisation parameters, see \citep{2021MNRAS.505.3505K}.
    \label{fig:QFP_phases}
    }
\end{figure}

As discussed in Sec.~\ref{sec:qfp}, the observed properties of guided QFP waves, such as modulation of the instantaneous amplitude and oscillation period, are highly influenced by the perpendicular profile of the plasma non-uniformity through the dispersion of fast magnetoacoustic waves in a waveguide. Thus, the time history of an impulsively excited QFP wave train in plasma slabs with a sufficiently smooth cross-field density profile resembles a \lq\lq crazy tadpole\rq\rq\ shape with a long, almost monochromatic tail propagating ahead and followed by a compact broadband head in the wavelet spectrum (see e.g. Fig. 3 in \citep{2004MNRAS.349..705N} and Fig. 4 in \citep{2003A&A...406..709K} for modelling and observational illustrations, respectively). In cylindrical waveguides, QFP wave trains  were found numerically to manifest in the wavelet spectrum as tadpoles propagating both tail-first as in the slab geometry \citep{2017ApJ...836....1Y} and head-first, i.e. being not crazy \citep{2015ApJ...814..135S}. Clarification of whether it is a physical or numerical effect is currently required for its use for seismology. The comparison between the QFP wave properties in plain and cylindrical geometries was also performed in \citep{2018ApJ...855...53L}.
Likewise, QFP wave trains' time signatures are expected to differ for waveguides with smoother and steeper density profiles. Thus, in plasma slabs with a steep density profile, the group speed of guided fast magnetoacoustic waves has a dip \citep{1995SoPh..159..399N} (or multiple dips, \citep{2016ApJ...833...51Y}), which results in the gradual transformation of a tadpole shape into a \lq\lq boomerang\rq\rq\ shape with two distinct arms at shorter and longer periods in the wavelet spectrum as the QFP wave train propagates along the waveguide (\citep{2021MNRAS.505.3505K}, see Fig.~\ref{fig:QFP_phases}). In the time domain, three distinct phases of such a boomerang-shaped wave train can be distinguished which are a long and narrowband {quasi-periodic phase} (I), intrinsically broadband and most energetic {peloton} phase (II) in which essentially short and long periods co-exist (i.e. propagate at the same group speed), and a short-lived periodic (Airy) phase (III). The apparent duration of the quasi-periodic and peloton phases in time, as potential observable, changes with the Alfv\'en speed ratio outside and inside the waveguide, $C_\mathrm{Ae}/C_\mathrm{Ai}$ and steepness of the cross-field density profile (see Fig.~3 in \citep{2021MNRAS.505.3505K} and Fig.~\ref{fig:QFP_phases}), resulting in the corresponding change in the wavelet spectral shape.

Observations in the radio band with intrinsically high temporal resolution seem to be most suitable for studying time signatures of QFP wave trains and using them for seismological diagnostics. For example, a crazy tadpole wavelet structure propagating upwards in the corona was detected in the radio burst observed with the Ondrejov radio spectrograph \citep{2011SoPh..273..393M}. The crazy tadpole shape suggests the observed event may be a QFP wave train propagating along a plain coronal plasma non-uniformity. In turn, the very nature of the observed radio emission requires the presence of a current sheet for charged particle acceleration. On this basis, the observed event can be interpreted as a QFP wave propagating along the plasma slab with a current sheet, as considered in e.g. \citep{2012A&A...537A..46J}. Moreover, the development of a second arm with height in the wavelet spectrum observed by \citep{2011SoPh..273..393M} is consistent with the evolution of QFP wave trains from tadpoles to boomerangs in steep waveguides, predicted theoretically by \citep{2021MNRAS.505.3505K}.
{More examples of the seismological analysis with QFP wave trains, e.g. for estimating the local direction of the coronal magnetic field, can be found in a series of works \cite{2012ApJ...753...52L, 2018ApJ...860...54O, 2021ApJ...908L..37M}.}

In summary, the phenomenon of dispersively evolving guided QFP wave trains potentially allows for probing such properties of the coronal plasma as:
\begin{itemize}
    \item Cylindrical or plain geometries of the wave-hosting plasma structure.
    \item Steepness of the perpendicular density profile.
    \item Density contrast inside the wave-hosting plasma structure and the surrounding corona.
   \item Apparent local direction of the coronal magnetic field.
\end{itemize}

However, the current use of guided dispersively evolving QFP waves for coronal seismology remains rather qualitative and would benefit from both theoretical and observational developments.

\subsection{Mapping the Alfv\'en speed by the global coronal waves}
\label{sec:globseis}

The propagation of the fast component of the global coronal wave is determined by the 3D structure of the fast speed, which, in turn, in the low-$\beta$ coronal plasma, is about the Alfv\'en speed.  Commonly, the seismological information comes from a visual inspection of EUV movies. The deformation of the wave front by refraction, i.e., the turn of the local wave vector indicated the direction in which the local fast speed decreases. Furthermore, the absolute value of the local magnetic field could be roughly estimated by the local phase speed $V_\mathrm{ph}$ as
\begin{equation}
\label{Eq:eitsei}
B_0 \approx \sqrt{\mu_0 \rho_0 (V_\mathrm{ph}^2 - C_\mathrm{s})},
\end{equation}
provided the phase speed coincides with the fast speed \cite{2013ApJ...776...55K}, and the local plasma density $\rho_0$ and the sound speed $C_\mathrm{s}$ could be reliably estimated. As the next step one can employ the short-wavelength Wentzel--Kramer--Brillouin approximation, taking that the local phase speed is
\begin{equation}
\label{Eq:fastsei}
V_\mathrm{ph}^2 = \frac{1}{2}\Big( C_\mathrm{A}^2 + C_\mathrm{s}^2 + \sqrt{(C_\mathrm{A}^2+C_\mathrm{s}^2)^2 - 4 C_\mathrm{A}^2C_\mathrm{s}^2\cos^2\phi} \Big),
\end{equation}
where $\phi$ is the angle between the wave vector and magnetic field \cite{2017ApJ...836..246K}. 

{As the next step, the global geometry of the coronal magnetic field could be determined. A comprehensive review of several pioneering studies of this approach, e.g., \cite{2011ApJ...730..122W, 2013SoPh..288..567L, 2013ApJ...776...55K} is given in \cite{2014SoPh..289.3233L}. A recent work which addressed a 3D model replicating a global coronal wave event, followed by seismology and comparison with model values, confirmed that global coronal waves can be used to probe the solar corona \cite{2021ApJ...911..118D}.}

\section{Future perspectives}
\label{sec:con}

MHD seismology of the solar corona is a rapidly developing mature branch of modern plasma astrophysics.
The anticipated next steps which are expected to advance it further could be summarised as follows.
\begin{itemize}

\item Ubiquitous decayless oscillations of coronal loops provide us with information about physical conditions in coronal active regions. The search for solar flare precursors in the variation of those parameters is of great interest for flare and mass ejection forecasting. Similar efforts are carried out with the use of apparent pre-flare oscillatory variations of the soft-X-ray and microwave emission, e.g., \cite{2016ApJ...833..206T, 2022Ge&Ae..62..895A, 2023Ge&Ae..63..513Z}. 
Furthermore, it has been demonstrated that kink oscillations can probe the magnetic field twist and the loop's sigmoidity \cite{2020ApJ...894L..23M}. It opens up interesting perspectives for the estimation of the non-potential (\lq\lq free\rq\rq) magnetic energy in pre-flaring active regions.  

\item Seismological applications of global coronal waves require the determination of whether the waves are in general spherical or cylindrical. In the former case, the wave propagate, in general, in all directions from the epicentre, i.e., a part of the wave energy leaves the corona for the solar wind. In the latter case, the refraction caused by the height non-uniformity of the fast speed, returns the waves to the corona, and the wave propagation is, in general, confined to the surface of the Sun. One can observationally distinguish between these two scenarios by the dependence of the wave amplitude on the distance. In a spherical wave the amplitude is proportional to the reciprocal of the distance, while in the cylindrical case, of its square {root}. Obviously, this estimation requires averaging the amplitude over a large segment of the wave front, to suppress the local deformations of the front by local plasma non-uniformities. The theoretical component of the seismological inversion needs to be based on solving the 3D fast wave equation with a non-uniform fast speed. An intrinsic difficulty will be accounting for the linear coupling of various MHD waves. Furthermore, the effect of the fine structuring of the coronal plasma on the evolution of the wave front needs to be revealed.  

\item An important element of the plasma diagnostics by MHD waves is to identify clearly the relationship between the plasma parameters perturbed by the wave and observables. In some cases this relationship is non-trivial, for example, variations of the coronal emission intensity could be caused by the variation of the plasma density or the column depth of the emitting plasma volume. If the former scenario indicates the compressive nature of the wave, the latter may be caused by a weakly compressive or even incompressive wave. This demonstrates the importance of forward modelling of the observational manifestation of various wave modes in various plasma structures, detected with specific instruments. The series of pioneering studies addressing this issue \cite{2013A&A...555A..74A, 2016FrASS...3....4V, 2016ApJ...820...13M, 2023MNRAS.520.4147K} requires further development.   

\item An interesting seismological opportunity is offered by the recently established dispersion of coronal slow magnetoacoustic waves, associated with thermal misbalance \citep{2019PhPl...26h2113Z, 2021SoPh..296..122B}. Characteristic signatures of dispersively formed wave trains carry information about the plasma heating. Similarly, properties of standing wave should be also affected by the non-adiabatic processes.   

\item A promising avenue is to utilise the complimentarity of independent techniques for the determination of the coronal magnetic field. In particular, seismological estimations could be used as reference points (or, rather, magnetic flux tubes) in the extrapolation of the field by photospheric sources during time intervals of the quiet solar activity. The omnipresent decayless kink oscillations and propagating slow waves seem to suit well this purpose.  

\item A major step forward in coronal seismology could be made with the use of different wave modes of the same plasma structure, as it increases the number of independent observables. The estimation of the plasma parameters variations along a coronal loop by multiple parallel harmonics of the kink mode have been briefly mentioned in Section~\ref{sec:magseis}. Furthermore, if slow and QFP waves are detected to propagate along the same path, the observed difference in phase speeds gives us information about the plasma $\beta$.     

\item Another interesting future direction is to transfer the MHD seismology to the diagnostics of stellar coronal plasma. It would be done in two approaches: the first one is to directly transfer the MHD seismology techniques to the stars that could be directly imaged by satellite or ground-based giant telescopes; the second one is to develop a reliable mode in QPP of solar flare and applied to stars that exhibit QPP during stellar flares. The QPPs of stellar flares would provide extra dimensions that stellar flares could be studied. Some initial ground for the latter approach has been created by revealing analogies between QPP in solar and stellar flares, e.g., \cite{2016ApJ...830..110C, 2018MNRAS.475.2842D}. 

\item In standing and sloshing slow oscillations, global coronal waves and decaying kink oscillations, the relative amplitude exceeds 10--20\%. Such strong disturbances are subject to nonlinear effects. Theoretical modelling of nonlinear processes, forward modelling of their manifestation in observational data and detection of the predicted signatures, and designing seismological techniques is another perspective research avenue. 

\item So far, the preliminary detection of coronal wave and oscillatory processes is usually based on the visual inspection of data sets of interest. This initial step of coronal seismology would highly benefit from the use of machine learning techniques which are designed to recognise characteristic patterns of known types of coronal wave phenomena.  
\end{itemize}

The development and application  of those novel MHD seismology techniques may follow the examples presented in this tutorial. 

\backmatter

\bmhead{Acknowledgments}
The following fundings are gratefully acknowledged:
China Scholarship Council-University of Warwick joint scholarship (S.Z.), the Latvian Council of Science Project No. lzp2022/1-0017 (D.Y.K. and V.M.N.), and the STFC consolidated grant ST/X000915/1 (D.Y.K.). D.Y. is supported by the National Natural Science Foundation of China (NSFC,12173012,12111530078), the Guangdong Natural Science Funds for Distinguished Young Scholar (2023B1515020049), the Shenzhen Technology Project (GXWD20201230155427003-20200804151658001), and the Shenzhen Key Laboratory Launching Project (No. ZDSYS20210702140800001). The data is used courtesy of the SDO/AIA team.

\bmhead{Competing Interests}
The authors declare that they have no conflicts of interests.

\bibliography{MRPP}
\end{document}